\documentclass[twocolumn]{aastex63}
\usepackage{CJKutf8}

\usepackage{verbatim}

\shorttitle{dust rings}
\shortauthors{Long et al.}

\graphicspath{{./}{figures/}}

\begin{document}
\begin{CJK*}{UTF8}{gbsn}

\title{Dual-Wavelength ALMA Observations of Dust Rings in Protoplanetary Disks}

\correspondingauthor{Feng Long}
\email{feng.long@cfa.harvard.edu}

\author{Feng Long(龙凤)}
\affiliation{Harvard-Smithsonian Center for Astrophysics, 60 Garden Street, Cambridge, MA 02138, USA}
\affiliation{Kavli Institute for Astronomy and Astrophysics, Peking University, Beijing 100871, China}

\author{Paola Pinilla}
\affiliation{Max-Planck-Institut f\"{u}r Astronomie, K\"{o}nigstuhl 17, 69117, Heidelberg, Germany}

\author{Gregory J. Herczeg(沈雷歌)}
\affiliation{Kavli Institute for Astronomy and Astrophysics, Peking University, Beijing 100871, China}

\author{Sean M. Andrews}
\affiliation{Harvard-Smithsonian Center for Astrophysics, 60 Garden Street, Cambridge, MA 02138, USA}

\author{Daniel Harsono}
\affiliation{Leiden Observatory, Leiden University, P.O. box 9513, 2300 RA Leiden, The Netherlands}
\affiliation{Institute of Astronomy and Astrophysics, Academia Sinica, No. 1, Sec. 4 Roosevelt Rd., Taipei 10617, Taiwan, Republic of China}

\author{Doug Johnstone}
\affiliation{NRC Herzberg Astronomy and Astrophysics, 5071 West Saanich Road, Victoria, BC,   V9E 2E7, Canada}
\affiliation{Department of Physics and Astronomy, University of Victoria, Victoria, BC, V8P 5C2, Canada}

\author{Enrico Ragusa}
\affiliation{School of Physics and Astronomy, University of Leicester, Leicester LE1 7RH, UK}

\author{Ilaria Pascucci}
\affiliation{Lunar and Planetary Laboratory, University of Arizona, Tucson, AZ 85721, USA}
\affiliation{Earths in Other Solar Systems Team, NASA Nexus for Exoplanet System Science, USA}

\author{David J. Wilner}
\affiliation{Harvard-Smithsonian Center for Astrophysics, 60 Garden Street, Cambridge, MA 02138, USA}

\author{Nathan Hendler}
\affiliation{Lunar and Planetary Laboratory, University of Arizona, Tucson, AZ 85721, USA}

\author{Jeff Jennings}
\affiliation{Institute of Astronomy, University of Cambridge, Madingley Road, Cambridge CB3 0HA, UK}

\author{Yao Liu}
\affiliation{Purple Mountain Observatory \& Key Laboratory for Radio Astronomy, Chinese Academy of Sciences, No.10 Yuanhua Road, Nanjing 210033, China}
\affiliation{Max-Planck-Institut f\"{u}r Extraterrestrische Physik, Giessenbachstrasse 1, 85748, Garching, Germany}

\author{Giuseppe Lodato}
\affiliation{Dipartimento di Fisica, Universita Degli Studi di Milano, Via Celoria, 16, I-20133 Milano, Italy}

\author{Francois Menard}
\affiliation{Univ. Grenoble Alpes, CNRS, IPAG, F-38000 Grenoble, France}


\author{Gerrit van de Plas}
\affiliation{Univ. Grenoble Alpes, CNRS, IPAG, F-38000 Grenoble, France}

\author{Giovanni Dipierro}
\affiliation{School of Physics and Astronomy, University of Leicester, Leicester LE1 7RH, UK}


\begin{abstract}
We present new Atacama Large Millimeter/submillimeter Array (ALMA) observations for three protoplanetary disks in Taurus at 2.9\,mm and comparisons with previous 1.3\,mm data both at an angular resolution of $\sim0\farcs1$ (15\,au for the distance of Taurus). In the single-ring disk DS Tau, double-ring disk GO Tau, and multiple-ring disk DL Tau, the same rings are detected at both wavelengths, with radial locations spanning from 50 to 120\,au. To quantify the dust emission morphology, the observed visibilities are modeled with a parametric prescription for the radial intensity profile. The disk outer radii, taken as 95\% of the total flux encircled in the model intensity profiles, are consistent at both wavelengths for the three disks. Dust evolution models show that dust trapping in local pressure maxima in the outer disk could explain the observed patterns. Dust rings are mostly unresolved. The marginally resolved ring in DS Tau shows a tentatively narrower ring at the longer wavelength, an observational feature expected from efficient dust trapping. The spectral index ($\alpha_{\rm mm}$) increases outward and exhibits local minima that correspond to the peaks of dust rings, indicative of the changes in grain properties across the disks. The low optical depths ($\tau\sim$0.1--0.2 at 2.9\,mm and 0.2--0.4 at 1.3\,mm) in the dust rings suggest that grains in the rings may have grown to millimeter sizes.
The ubiquitous dust rings in protoplanetary disks modify the overall dynamics and evolution of dust grains, likely paving the way towards the new generation of planet formation.  
\end{abstract}


\section{Introduction} \label{sec:intro}


In the standard core-accretion scenario of planet formation, dust grains have to grow from micron-sized solids to millimeter/centimeter-sized pebbles then to kilometer-sized planetesimals, which eventually build up terrestrial planets and the cores of giant planets. This transformation in grain sizes is dramatic and challenging in a few Myr timescale.  Observations of protoplanetary disks at (sub-)millimeter wavelengths are thus essential to probe the first steps of planet formation (see review by \citealt{Testi2014}).

In a disk with a smooth gas distribution, dust particles of millimeter or centimeter sizes at disk outer regions suffer from severe aerodynamic drag, which pushes them inward \citep{Weidenschilling1977}. Large grains should therefore be largely depleted in the outer disks ($\gtrsim$20\,au) within 1\,Myr \citep{Brauer2008, Birnstiel2010}. In contrast, millimeter observations reveal many disks extending to hundreds of au in radius after a few Myr evolution \citep[e.g.,][]{Andrews2007}. One natural solution for this contradiction between observations and theoretical predictions involves local pressure bumps in disks (i.e. gas distribution is not smooth), which halt the inward drift, trap dust particles, and retain large grains at wide radial distances \citep{Whipple1972,Nakagawa1986,Pinilla2012_bump}.

Recent high-resolution continuum observations from the Atacama Large Millimeter/submillimeter Array (ALMA), show that distributions of millimeter-sized grains in protoplanetary disks are highly structured, often seen as axisymmetric gaps and rings \citep[e.g.,][]{Isella2016,Cieza2017,Fedele2018, vanTerwisga2018,Clarke2018,Andrews2018_dsharp, Long2018, vandermarel2019}. In a survey of 32 Taurus disks at $\sim$0$\farcs$1 resolution with ALMA, \citet{Long2019} found that disks with dust radii larger than 55\,au (measured from 1.3\,mm continuum emission) all host substructures. The presence of millimeter dust grains at large radii and the structured nature of the dusty disk provide observational support for dust trapping as the solution to the radial drift problem in disks, and are usually attributed to the dynamical interaction between young planets and the disk \citep[e.g.,][]{Pinilla2012_Planet,Dipierro2015, Dong2017}. However, in this scenario the formation process of the first generation of planets that would be responsible for the pressure bumps is still unclear. Other origins of pressure bumps, including zonal flows, gradients of disk viscosity, and the secular gravitational instability are also widely discussed in the literature \citep[e.g.,][]{Youdin2011,Johansen2009,Takahashi2014, Flock2015}.

Observational evidence for dust trapping in disk pressure maxima could be investigated with multiple approaches. For example, if the dust ring is narrower than the gas pressure bump, dust trapping must have occurred. \citet{Dullemond2018} applied this idea to the DSHARP (Disk Substructures at High Angular Resolution Project) sample at 2--3\,au resolution by comparing the measured width of dust rings with the estimated gas pressure scale height, and reported strong dust trapping in some cases. This method requires very high spatial resolution continuum observations and appropriate estimate of gas pressure profile, which is also observationally challenging (but see \citealt{Teague2018}). Dust trapping models also predict that larger grains accumulate more efficiently in the pressure maxima than smaller size particles, thus forming a narrower distribution when trapped \citep{Birnstiel2013, Pinilla2015}. Disk observations at different wavelengths, tracing grains of different sizes, would be an ideal test. In some transition disks, dust cavities at millimeter wavelengths are wider than what have been seen from near-infrared scatter light,  as expected from dust trapping models with massive planets \citep{Hendler2018,Villenave2019}. The comparison of dust rings at 0.45, 1.30, and 2.75\,mm in the SR 24S transition disk is consistent with dust trapping models \citep{Pinilla2019}, while comparison at only short wavelengths (0.45, 0.88 and/or 1.30\,mm) sometimes leads to ambiguous interpretations \citep{Pinilla2015}, mainly due to high optical depth. Observations at longer wavelengths, with the benefit of lower optical depth, would therefore be crucial to test particle trapping, and are still largely absent for the recently discovered multi-ring disks.

Multi-wavelength observations are also essential to assess the dust grain properties. Evidence of the presence of large grains (millimeter-sized) in protoplanetary disks are provided by the spatially-integrated measurements of the spectral index from sub-millimeter to centimeter wavelength range \citep{Andrews2005, Ricci2010_oph, Ricci2010}. Spatially-resolved observations make the measurements of radial variations of grain properties possible. For instance, \cite{Perez2012} and \cite{Tazzari2016} found lower spectral index (enhanced grain growth) in the inner disk compared to the outer disk. More striking variations are witnessed across the dust gaps and rings, seen as lower spectral index in the bright rings and higher values in the depleted gaps \citep{alma2015,Tsukagoshi2016,Huang2018_CO,Macias2019,Carrasco2019,Huang2020}. The high density, as well as high dust-to-gas ratio of dust rings, could facilitate rapid planetesimal formation thus serving as promising sites for planet formation. The observed low spectral index could be a hint of grain growth in dust concentrations, but could also be the result of large optical depth \citep{Pinte2016, Dent2019}. It is therefore necessary to explore the radial change of grain properties with optically thin dust rings.

In this paper, we select three disks (DS Tau, GO Tau, and DL Tau) with optically thin rings identified from our previous 1.3\,mm survey at 0$\farcs$1 resolution (or equivalently, 15\,au resolution) from \citet{Long2018}. 
They represent disks with single ring, double rings, and complex rings. Here, we present the analysis of the three disks at both 1.3 and 2.9\,mm, to characterize the dust distributions for different grain sizes. This comparison aims to test the presence of dust traps and also provide insights for grain property changes, to better understand the role of dust substructures in planet formation process. In Sect.~\ref{sec:obs}, we present the ALMA Band\,3 (2.9\,mm) observations for the three disks. The morphology comparison at two wavelengths, the derived disk dust radius and dust ring properties from visibility fitting, as well as the mapped spectral index profiles are presented in Sect.~\ref{sec:obs-results}. We discuss our results from observations in the context of dust evolution models in Sect.~\ref{sec:discussion} and summarize our findings in Sect.~\ref{sec:sum}.

\begin{deluxetable*}{lcccccccrcc}
\tabletypesize{\scriptsize}
\tablecaption{Host Stellar Properties and Observation Results \label{tab:source_prop}}
\tablewidth{0pt}
\tablehead{
\colhead{Name} & \colhead{2MASS} & \colhead{D} & \colhead{SpTy} & \colhead{$T_{\rm eff}$} & \colhead{$L_*$} & \colhead{$M_*$} & \colhead{$t_*$} & \colhead{frequency} & \colhead{RMS noise} & \colhead{beam size} \\
\colhead{} & \colhead{} & \colhead{(pc)} & \colhead{} & \colhead{(K)} & \colhead{($L_\odot$)} &  \colhead{($M_\odot$)} & \colhead{(Myr)} & \colhead{(GHz)} & \colhead{($\mu$Jy beam$^{-1}$)} & \colhead{($''\times''$)} 
} 
\colnumbers
\startdata
    DS Tau & 04474859+2925112  &  159 &   M0.4 & 3792 &  0.25 & 0.58$^{+0.17}_{-0.13}$  &  4.80$^{+ 4.80}_{-2.30}$ & 225.5 (B6)  & 82.5  & 0.13$\times$0.09 \\ 
    		&				   &	      &		    &		&	    &					    &					     &	105 (B3)     & 16.5  & 0.13$\times$0.09 \\ 
    GO Tau & 04430309+2520187 &  144 &   M2.3 & 3516 &  0.21 & 0.36$^{+0.13}_{-0.09}$  &  2.20$^{+ 1.90}_{-1.10}$ &  225.5 (B6) &  58.5 & 0.13$\times$0.10 \\
        		&				   &	      &		    &		&	    &					    &					     &	105 (B3)     & 12.9  & 0.13$\times$0.10 \\ 
    DL Tau & 04333906+2520382  &  159 &   K5.5  & 4277 &  0.65 & 0.98$^{+0.84}_{-0.15}$  &  3.50$^{+ 2.80}_{-1.60}$ &  225.5 (B6) &  60.5 & 0.13$\times$0.10 \\
        		&				   &	      &		    &		&	    &					    &					     &	105 (B3)     & 12.8  & 0.13$\times$0.10 \\
\enddata
\tablecomments{The distance for individual stars is adopted from the Gaia DR2 parallax \citep{gaia2018}. Spectral type is adopted from \citet{herczeg2014} and stellar luminosity is calculated from J-band magnitude and updated to the new Gaia distance. Stellar mass and age are adopted from \citet{Long2019}. The last three columns correspond to the central frequency, noise level, and final smoothed synthesised beam FWHM from our ALMA observations.}
\end{deluxetable*}

\section{ALMA Band\,3 Observations} \label{sec:obs}
Our ALMA observations at Band\,3 for DS Tau, GO Tau, and DL Tau were taken between 2019 July 16 and July 28 (\#2018.1.00614.S, PI.\,Long). The array was configured to span baselines from 90\,m to $\sim$8.5\,km with 43--45 antennas, to achieve comparable angular resolution to our previous Band\,6 data (with baselines from 21\,m to 3.6\,km). Three spectral windows were set up for continuum observations, centered at 98, 100, and 112\,GHz, each with a bandwidth of 1.875\,GHz. The remaining window was split for targeting $^{13}$CO and C$^{18}$O with a channel width of $\sim$0.7\,km\,s$^{-1}$. The total on-source integration time were 66.5~min, 77.6~min, and 67.3~min for DS Tau, GO Tau, and DL Tau, respectively.

The data were calibrated by ALMA pipeline with the Common Astronomy Software Package (CASA), version 5.4.0. Further calibration and imaging were also performed with this same version. Bandpass and flux calibrations used observations of the quasar J0510+1800 for all executions. The gain calibrations used the quasar J0438+3004 for DS Tau and GO Tau, and the quasar J0426+2327 for DL Tau. The final continuum dataset was created by combining the three continuum spectral windows with the line-free channels in the line spectral window, and binned into 125\,MHz channels, resulting in an average frequency of 105\,GHz (2.9\,mm). We performed two rounds of phase-only self-calibration with solution intervals of 120s and 60s (image quality did not improve when reducing the interval) for GO Tau and DL Tau. Only 20--40\% improvements in peak signal-to-noise ratio were seen after self-calibration. DS Tau was too faint for self-calibration to improve the image quality.

The Band\,3 continuum images were obtained with the multi-term, multi-frequency synthesis algorithm (\texttt{mtmfs}) in \texttt{tclean} with \textit{nterms} = 2. In order to compare the dust emission morphology at Band\,3 with the previous Band\,6 data in the image plane, the final images at both bands were convolved into a common beam size using the \texttt{imsmooth} task. The choice of initial weighting parameters and the common beam sizes were based on a compromise between observational sensitivity, which ensures substructures were well detected, and angular resolution, which renders multiple disk components well separated. For DS Tau, we started with Briggs weighting with robust = -0.5 at both bands, resulting in beam sizes of $0\farcs11\times0\farcs06$ and $0\farcs12\times0\farcs08$ for Band\,3 and Band\,6 images, respectively. Both images were then convolved to reach a targeted beam size of $0\farcs13\times0\farcs09$. For GO Tau and DL Tau, the initial images from robust = 0.0, which have beam sizes of $0\farcs11\times0\farcs07$ and $\sim0\farcs12\times0\farcs10$ for Band\,3 and Band\,6 images, were smoothed to images with a beam size of $0\farcs13\times0\farcs10$. The new Band\,3 observations have slightly better angular resolution. The 1$\sigma$ noise levels measured in the signal-free regions are in the range of 12 to 16\,$\mu$Jy\,beam$^{-1}$ (see Table~\ref{tab:source_prop} for more details).

\begin{figure*}[!t]
\centering
    \includegraphics[width=0.32\textwidth]{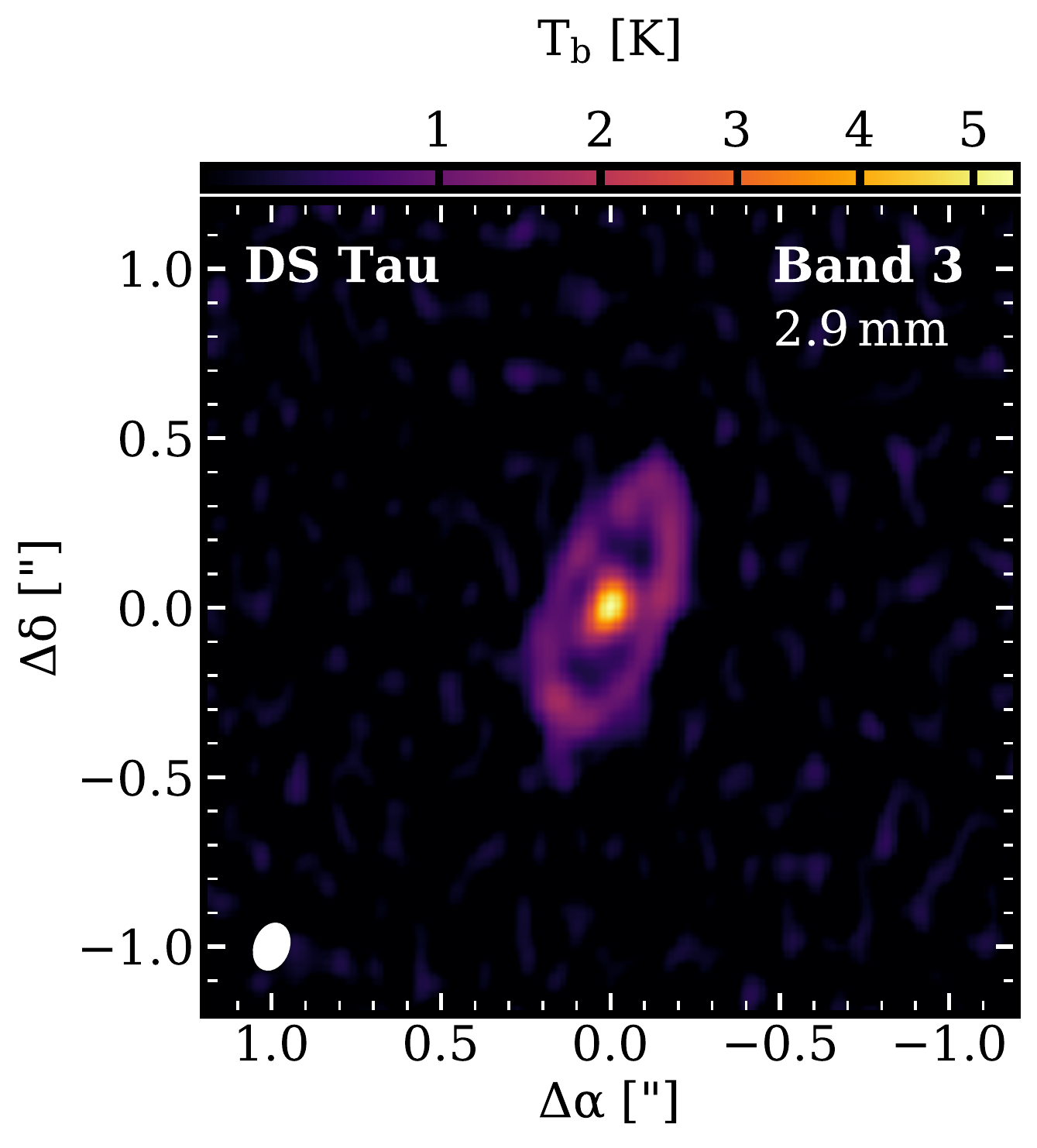} 
    \includegraphics[width=0.32\textwidth]{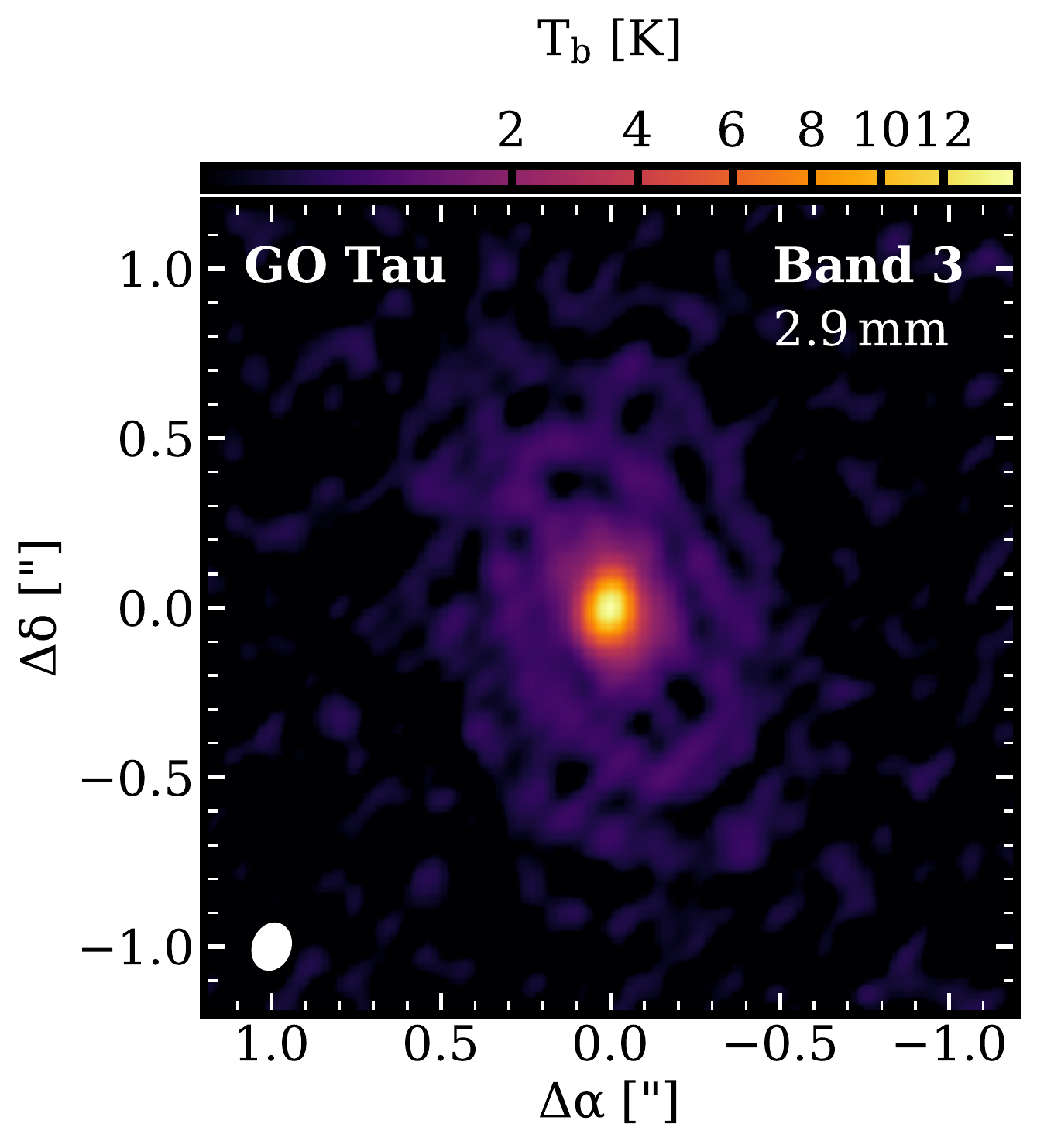} 
    \includegraphics[width=0.32\textwidth]{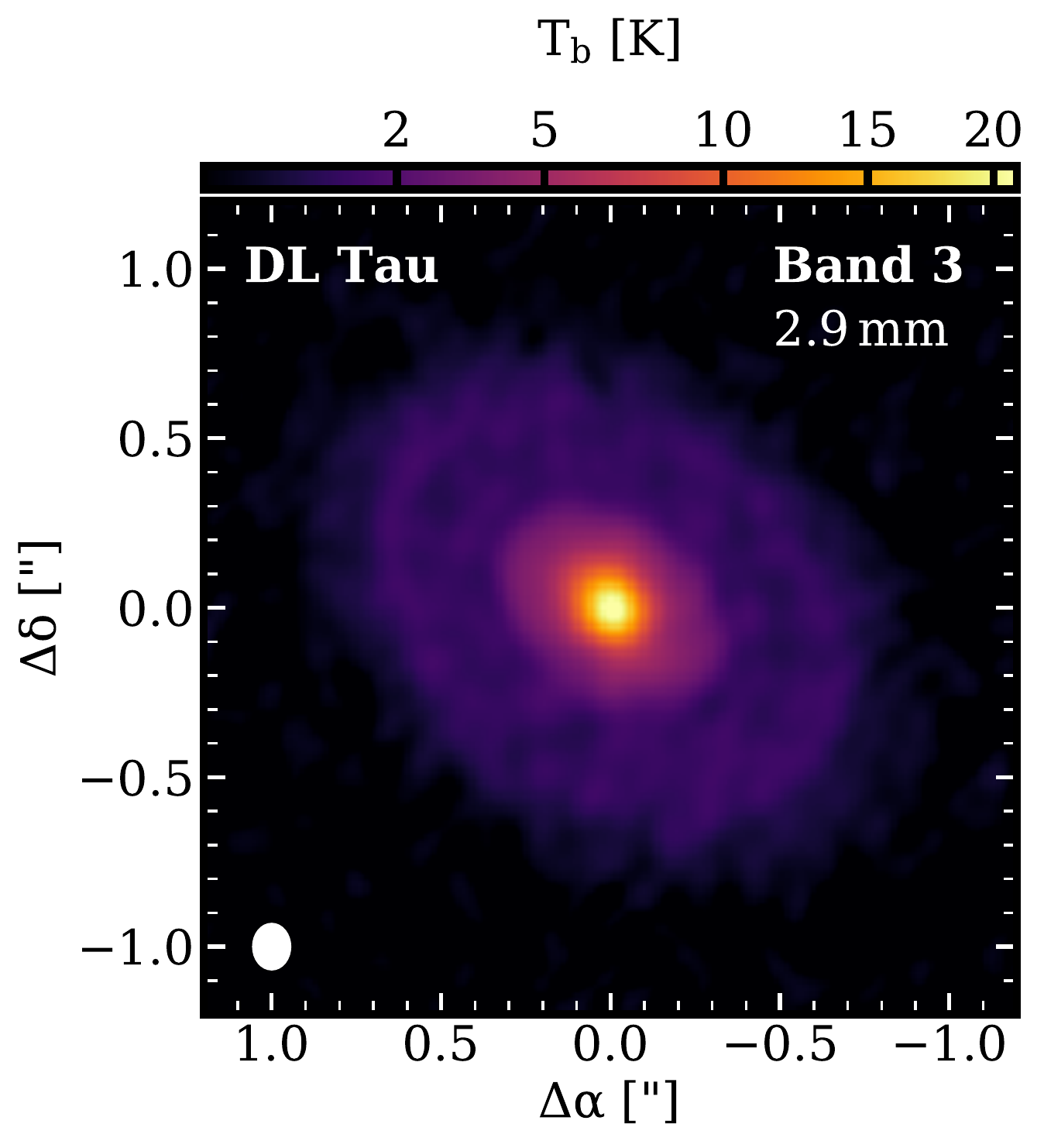} \\ 
    \includegraphics[width=0.32\textwidth]{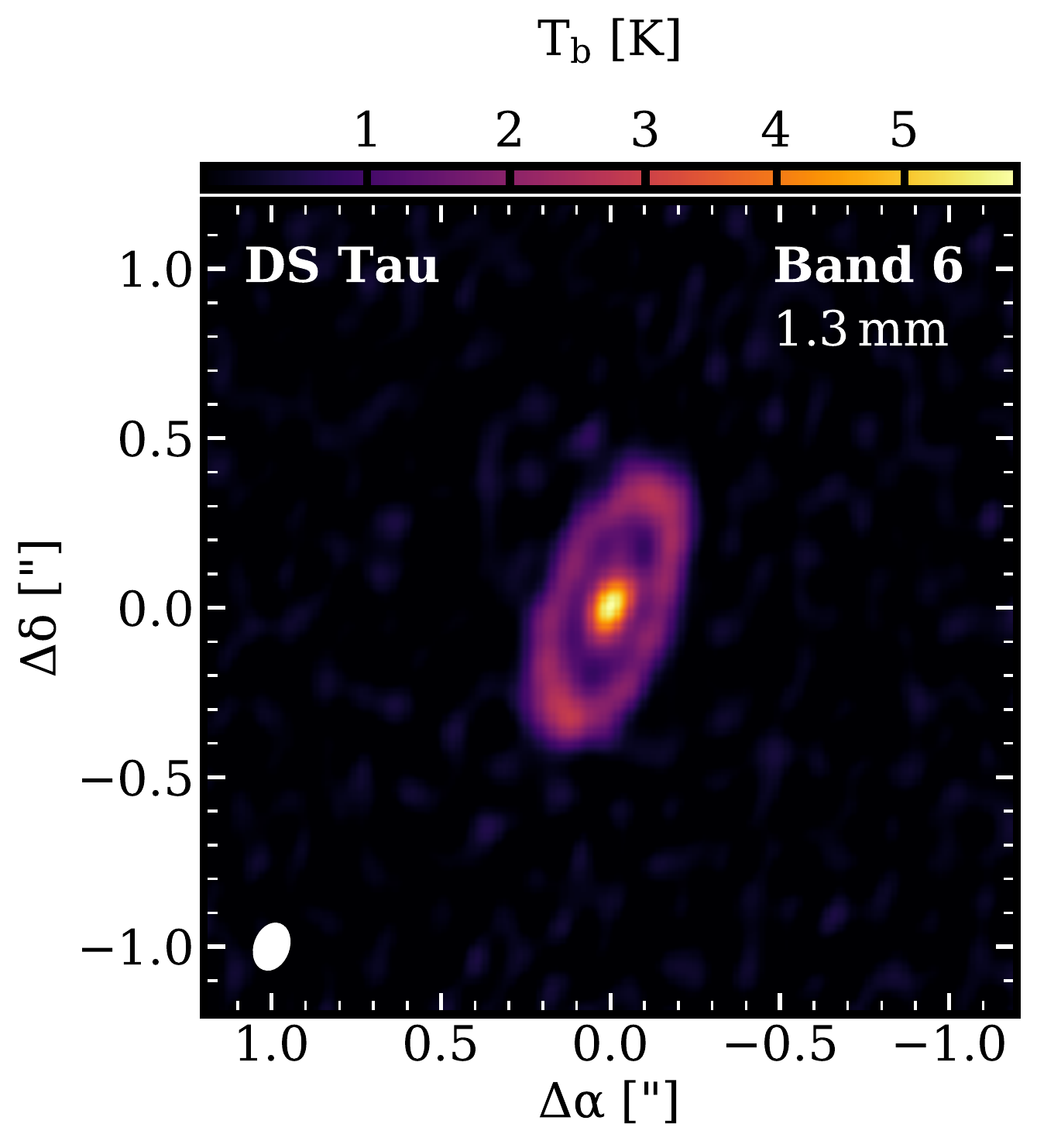}
    \includegraphics[width=0.32\textwidth]{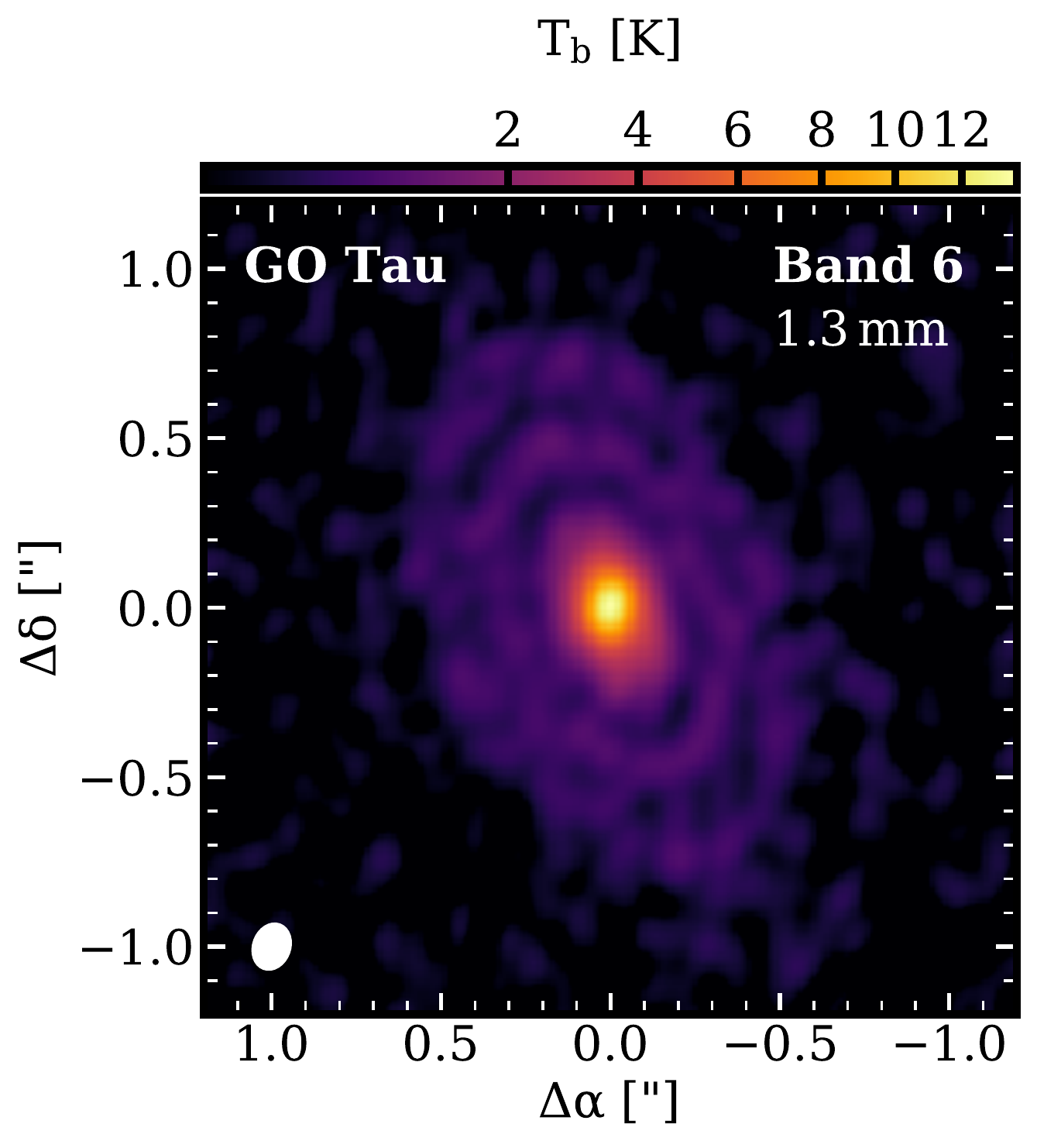}
    \includegraphics[width=0.32\textwidth]{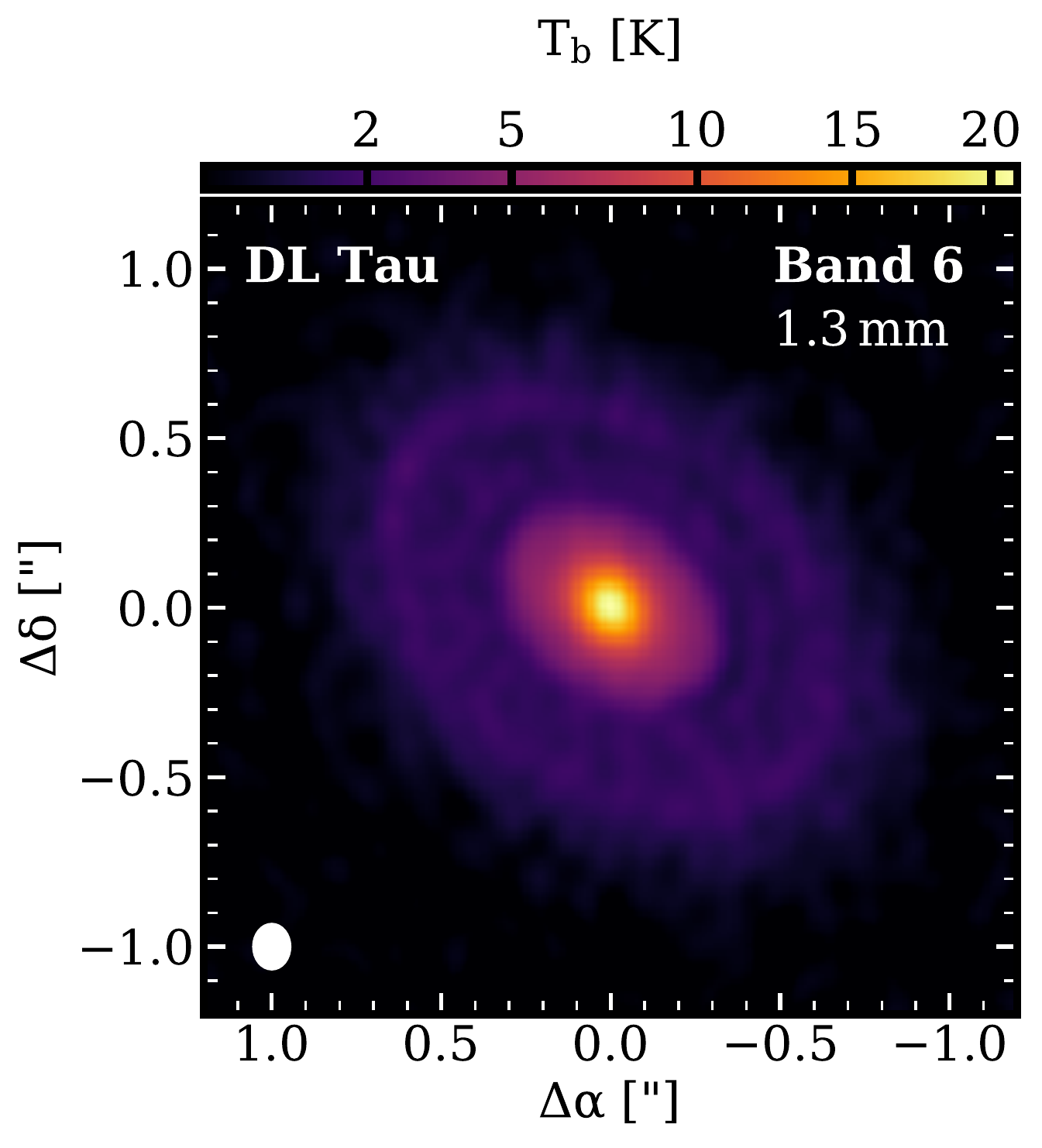}
    \caption{ALMA continuum images at 1.3\,mm (Band\,6, bottom panels) and 2.9\,mm (Band\,3, top panels) in brightness temperature calculated using Rayleigh-Jeans approximation, with identical synthesized beams for individual disks. The color scheme was applied with a power-law stretch to highlight the weak emission in dust rings. Dust emission at two wavelengths are very similar. \label{fig:cont_images} }
\end{figure*}

\begin{figure*}[!t]
\centering
    \includegraphics[width=0.325\textwidth]{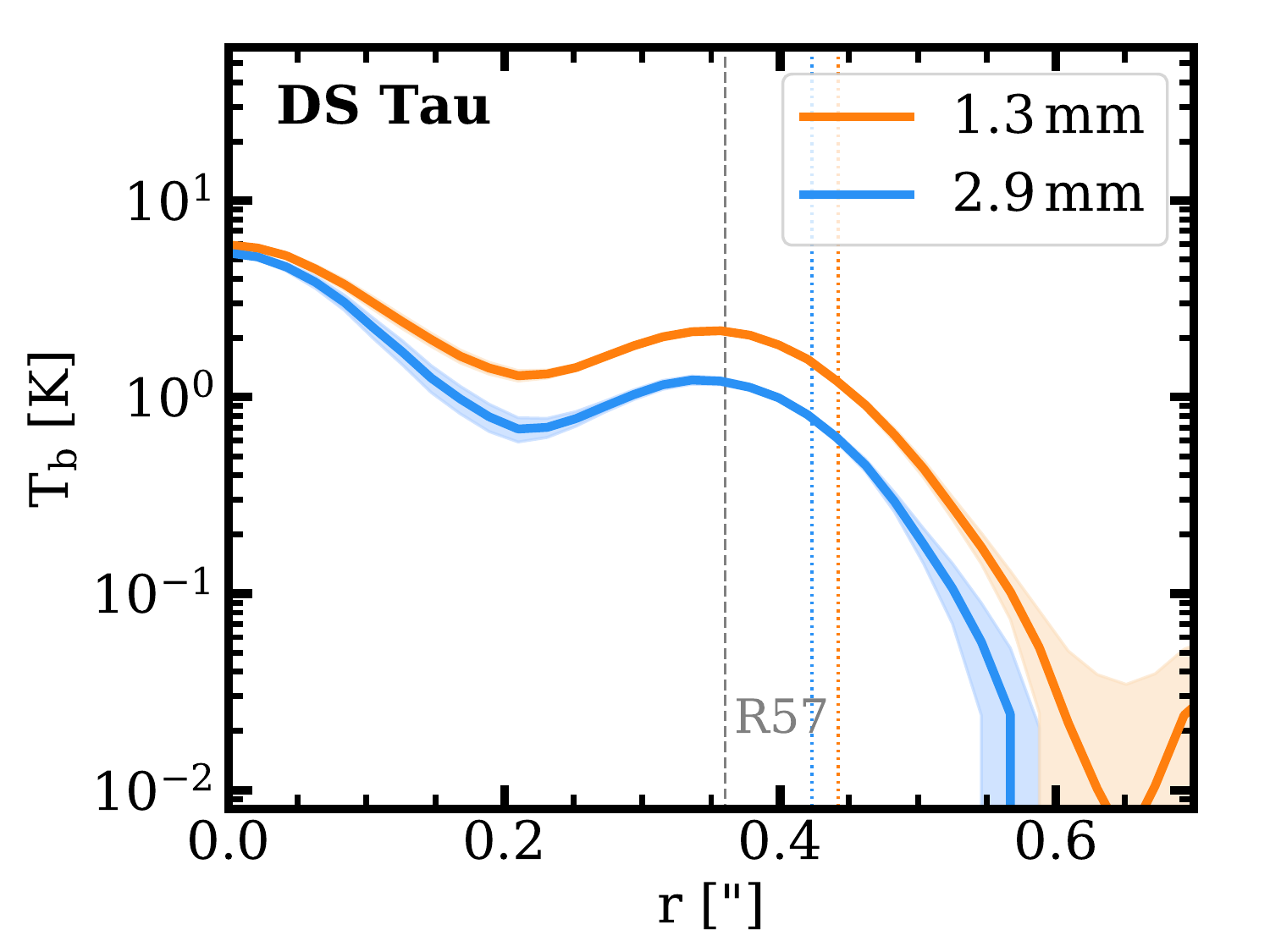} 
    \includegraphics[width=0.325\textwidth]{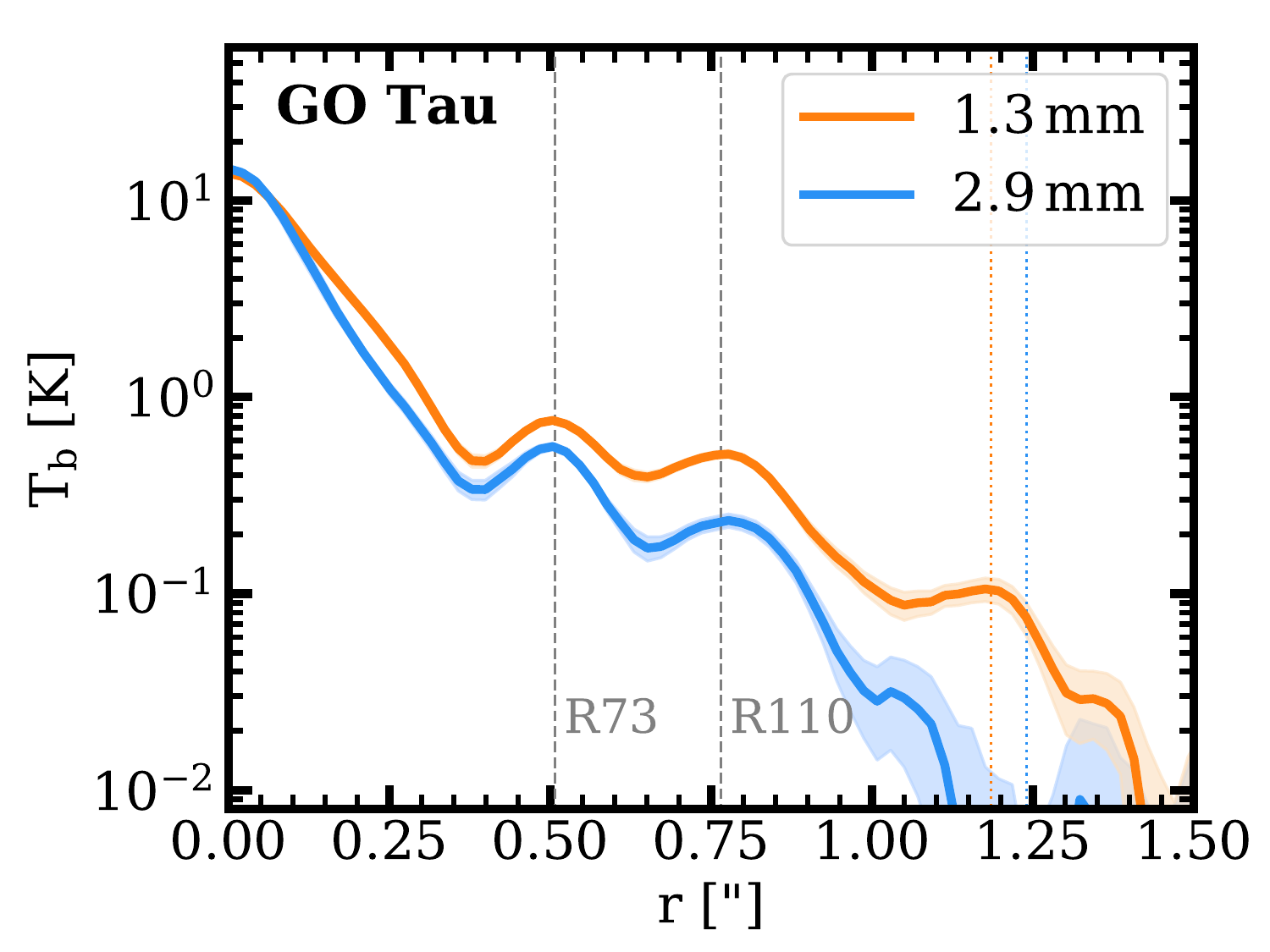}
    \includegraphics[width=0.325\textwidth]{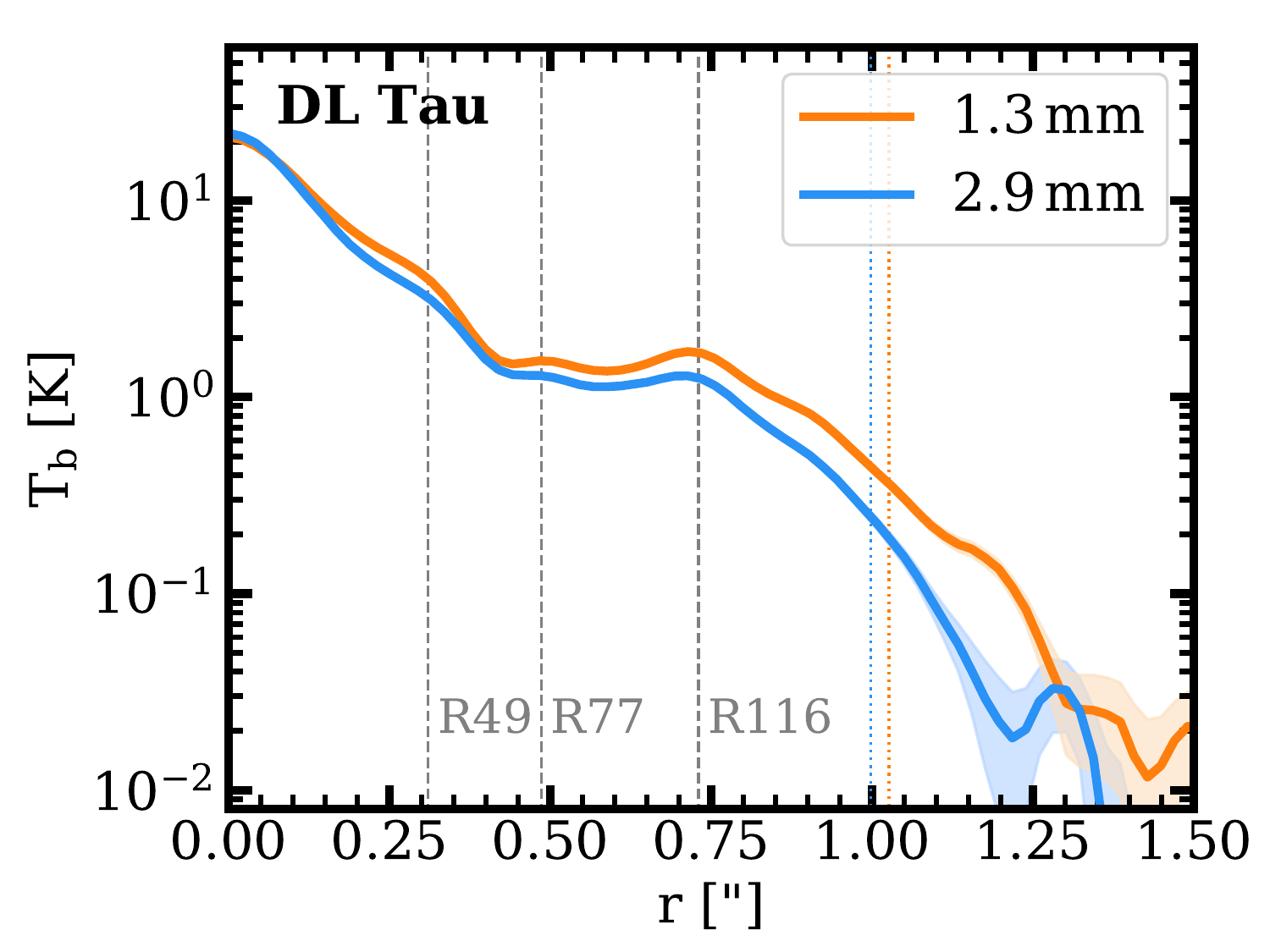} \\ 
    \caption{Deprojected and azimuthally averaged brightness temperature profiles in logarithmic scale, using the Rayleigh-Jeans approximation. The disk inclination and position angles used in the deprojection are adopted from \citet{Long2018}. Light shaded regions show the 1$\sigma$ scatter divided by the square root of the number of beams spanning the full azimuthal angle at each radial bin. Each prominent dust ring is highlighted with a dashed line and a label denoting the ring location. R$_{\rm eff,95\%}$ from model fittings are plotted as dotted lines with corresponding colors (see more discussions about disk radius comparison in Section~\ref{sec:radius}).} \label{fig:cont_rp}
\end{figure*}

\section{Results} \label{sec:obs-results}

\subsection{Disk Morphology} \label{sec:morphology}
The continuum images at 2.9\,mm (Band\,3) for our three disks are shown in Figure~\ref{fig:cont_images}.
The 1.3\,mm (Band\,6) images obtained from ALMA Cycle 4, are created with identical beam sizes for individual disks and displayed below for a direct comparison. The colorbar of this figure displays the brightness temperature, which is obtained assuming the Rayleigh-Jeans approximation. Dust emission at both wavelengths is detected towards similar radial extents with similar morphology in our sample. 
The azimuthally averaged radial intensity profiles from the deprojected images are shown in Figure~\ref{fig:cont_rp}.
The dust rings (``bright'' annuli) reported at 1.3\,mm images \citep{Long2018} are all detected in our new 2.9\,mm data at their corresponding locations, though with lower signal-to-noise ratios. DS Tau has an inner disk surrounded by one ring (R57\footnote{The number here represents the radial distance of the ring peak to the central star in au.}). GO Tau shows an inner disk plus two rings (R73 and R110), while the faint outer disk identified from 1.3\,mm radial profile (Figure~\ref{fig:cont_rp}) is mostly buried in the noise at 2.9\,mm.
DL Tau, the brightest disk in our sample, shows complex structures, including an emission bump (R49) well connected with the inner disk, a faint and very narrow ring (R77), and a slightly brighter ring (R116) embedded within some diffuse halo emission. 
We define the substructure depth as intensity ratio between the gap location (radius of the local minimum) and its associated outer ring location. High depths of 0.5--0.6 are seen in the R57 ring of DS Tau and the R73 ring of GO Tau. For dust rings in DL Tau, only $\sim$10\% contrasts in emission brightness are observed. The true gap-ring contrast should be larger than we estimate here due to beam smearing.
In all three systems, the inner disks, defined as the region inside the first local intensity minimum, are slightly more compact at 2.9\,mm than what is seen at 1.3\,mm (see Figure~\ref{fig:cont_rp}). 


To describe the morphology of millimeter continuum emission in our sample, we perform the disk modeling in the uv-plane. 
The observed visibilities are compared with synthetic visibilities of the model intensity profile. Given the similar morphologies for the disks at 1.3 and 2.9\,mm, we adopt the same intensity profiles as \citet{Long2018}, which are reasonably good models for the dust emission at 1.3\,mm with less than 5$\sigma$ residuals. DS Tau is modeled with a central Gaussian Profile for the inner disk and a Gaussian function centered at the location of its ring peak, which is \begin{equation}
I(r) = F_0 \exp \left( -\frac{ r^{2} }{2  \sigma_{0} ^{2} } \right) +  F_1 \exp \left[ -\frac{ (r- R_{1} )^{2} }{2  \sigma_{1} ^{2} } \right]. 
\end{equation}
The inner disks of GO Tau and DL Tau are modeled with an exponentially tapered power-law, since the falloff in the edge of the inner disk is sharper than a Gaussian profile that results in significant symmetric residuals ($>$5--10$\sigma$, \citealt{Long2018}). The profile is then expressed as 
\begin{equation}
I(r) = F_0{\left(\frac{r}{r_c}\right)}^{-\gamma_1} \exp\left[-\left(\frac{r}{r_c}\right)^{\gamma_2}\right] + \sum_i F_{i} \exp\left[-\frac{(r- R_{i})^{2}}{2\sigma_{i}^{2}}\right],
\end{equation}
where the power-law index $\gamma_1$ and taper index $\gamma_2$ describe the emission gradient of the inner disk. The number of ring components for each disk is counted by emission bumps in the radial profile and adjusted to account for the faint outer disk. Following \citet{Long2018}, we choose one ring for DS Tau, two rings plus one additional faint ring to model the outer disk for GO Tau, and three rings plus one broad ring component for the diffuse halo emission for DL Tau. The same functional forms, including same numbers of Gaussian rings, are used for the fitting of individual disks at both wavelengths.
With the defined model intensity profile, we generate synthetic visibilities using the \textit{Galario} code \citep{Tazzari2018}, sampled at the observed uv-space. The disk inclination and position angles, as well as phase center offsets, are all set as free parameters. Our fitting is then performed using \textit{emcee v3.0.1} \citep{ForemanMackey2013}, in which a Markov Chain Monte Carlo (MCMC) method is implemented to explore the free parameter space. The radial grid in the model is linearly spaced from 0$\farcs$0001 to 4$\farcs$, with steps of 0$\farcs$0005, much smaller than our beam size ($\sim0\farcs1$). We set uniform prior probability distributions for the free parameters as $p(logF_i) \in [6,11]$\,Jy/Sr, $p(\sigma_i) \in [0,0\farcs2]$\footnote{Prior of ring sigma for the additional Gaussian ring component for the diffuse outer disk is given as $[0, 0\farcs6]$.}, $p(\gamma_1$) $\in [0,2]$, $p(\gamma_2)$ $\in [0,20]$, and $p(r_{c}) \in [0,0\farcs4]$. Priors on the ring center locations are given as $[R_i-0\farcs05, R_i+0\farcs05]$, where $R_i$ is center location derived from 1.3\,mm data for individual dust rings.
Priors on disk inclination and position angles are centered at what identified before \citep{Long2018} with a range of $\pm$20$\degr$.  The free parameters are sampled with 100 walkers and 5000 steps for each walker. Given the typical autocorrelation time on the order of $10^2$, these steps are sufficient to reach convergence. The last 1000 steps are used to sample the posterior distribution. The adopted parameters are taken as the peaks of marginal posteriors, with uncertainties given by the 68\% confidence intervals (see Table~\ref{tab:uvmodel_prop}, see also Table~\ref{tab:dstau_uvmodel_prop},\ref{tab:gotau_uvmodel_prop},\ref{tab:dltau_uvmodel_prop} in the Appendix for the full list of parameters).


\begin{figure*}[!t]
\centering
    \includegraphics[width=0.98\textwidth]{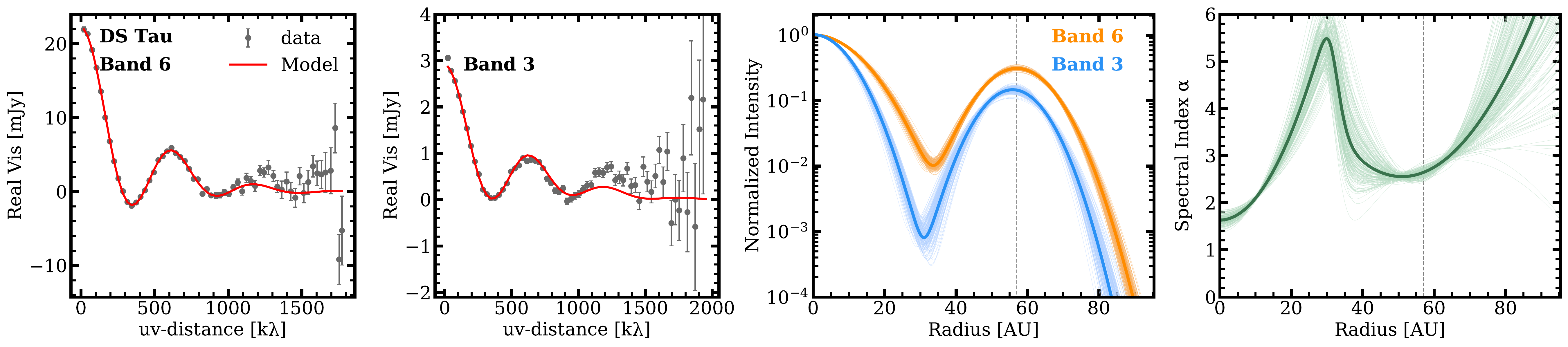} \\
    \includegraphics[width=0.98\textwidth]{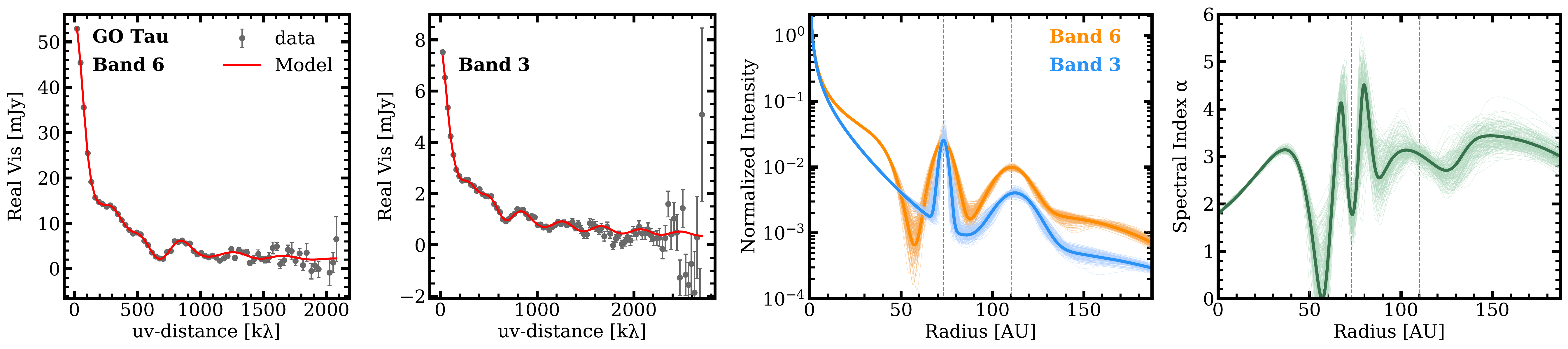} \\
    \includegraphics[width=0.98\textwidth]{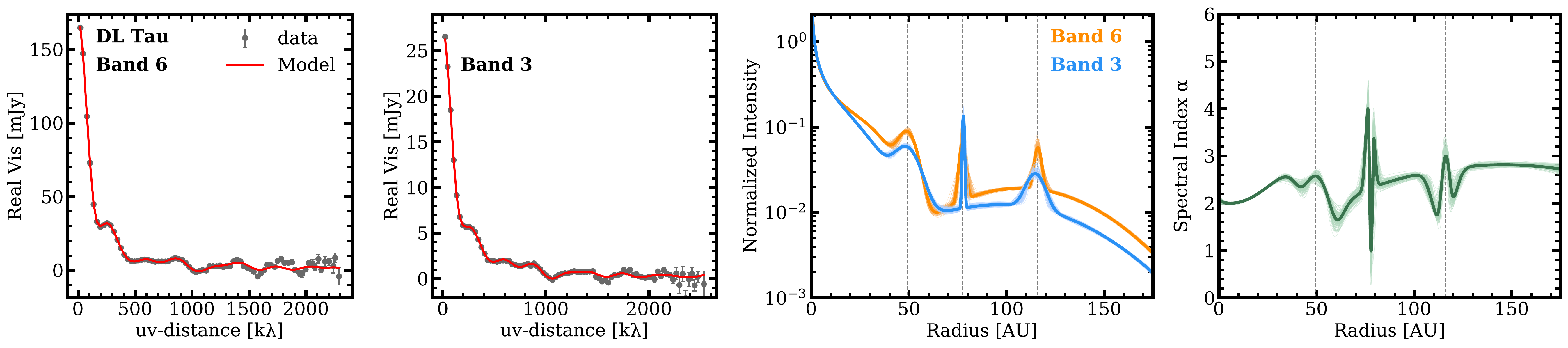} \\
    \caption{\textit{From left to right:} Comparison between the model and observed real part visibilities as a funcion of deprojected baseline length in 30\,k$\lambda$ bins; the adopted intensity profile model normalized to the intensity value at 0$\farcs$01, with 200 randomly selected chains overlaid; the model spectral index profile derived from model intensity profiles at the two wavelengths. \label{fig:uvmodel} }
\end{figure*}

Figure~\ref{fig:uvmodel} compares the adopted model visilibities to the binned real part of the data visilibities as a function of projected \textit{uv}-distance. The imaginary part of the data visilibities are flat around zero out to 1500\,k$\lambda$, consistent with our assumption of symmetric intensity models, and thus not shown. Our models match the overall structures in the visibility profiles reasonably well. As shown in Figure~\ref{fig:uvmodel-app} in the Appendix for the data, model, and residual map comparisons, the disk main structures (e.g., ring location and width) are well captured by the assumed models. However, 5--10$\sigma$ residuals are seen in the inner disk of GO Tau and DL Tau at Band\,3 (significant residuals are not seen for Band\,6 data), which indicate that our choice of the intensity models may not be the best form for dust emission of the inner disk at 2.9\,mm. This is also reflected in the imperfect match of data and model visibilities at long baselines (outward of 1200\,k$\lambda$, Figure~\ref{fig:uvmodel}), indicating the presence of small-scale features that are not captured by our models. The mismatch in DS Tau can be resolved by replacing the Gaussian profile (for the inner disk) with a tapered power-law function or a Nuker profile \citep{Tripathi2017}, which provide better fits to emission with a sharper transition than what is presented by a Gaussian profile. It is also possible that small-scale substructures are present inside 20\,au of the DS Tau disk \citep[see DSHARP,][]{Huang2018_ring}, however identifying them is challenging given the resolution of our data. Attempts at fitting with an additional Gaussian ring component in the inner disk failed to converge (after 10000 steps). We keep the simple Gaussian profile plus Gaussian ring model for DS Tau as this model describes the data reasonably well with only 3$\sigma$ residuals. We will discuss the effects of using different functions in the following when needed. The Band\,3 data have slightly finer resolution, and any substructures are easier to identify in the more optically thin long wavelength observations. Given the same intensity functions adopted at both wavelengths, the larger residuals in the Band\,3 modeling (especially for the inner disk of GO Tau and DL Tau) thus imply the presence of additional substructures, while the numbers and locations for these hidden features are difficult to quantify. The success of parametric fitting largely depends on the prior knowledge of the component numbers, thus the model for the inner disk of our three disks should be revised with further higher resolution observations. A 3$\sigma$ residual can be seen in the dust gap of DS Tau at 2.9\,mm, which has a tentative counterpart in the gap at 1.3\,mm (see Figure~\ref{fig:uvmodel-app} in the Appendix).


\begin{deluxetable*}{lccccccccccc}
\tabletypesize{\scriptsize}
\tablecaption{Source properties from visibility modeling \label{tab:uvmodel_prop}}
\tablewidth{0pt}
\tablehead{
\colhead{Name} & \colhead{Band} & \colhead{F$_{\nu}$} & \colhead{R$_{95\%}$} & \colhead{Incl} & \colhead{PA} & \colhead{R$_{1}$} & \colhead{$\sigma_1$} & \colhead{R$_{2}$} & \colhead{$\sigma_2$} & \colhead{R$_{3}$} & \colhead{$\sigma_3$} \\
\colhead{} & \colhead{} & \colhead{(mJy)} & \colhead{(au)} & \colhead{(deg)} & \colhead{(deg)} & \colhead{(au)} &  \colhead{(au)}  & \colhead{(au)} & \colhead{(au)} &  \colhead{(au)}  & \colhead{(au)}
} 
\colnumbers
\startdata
DS Tau & 1.3 mm & 22.15$_{-0.17}^{+0.24}$ & 70.30$_{-0.89}^{+0.58}$ & 65.23$_{-0.37}^{+0.30}$ & 159.74$_{-0.41}^{+0.32}$ &  57.01$_{-0.38}^{+0.40}$ &   8.14$_{-0.52}^{+0.36}$ &  &  &  &  \\
DS Tau & 2.9 mm &  2.90$_{-0.03}^{+0.02}$ & 67.29$_{-1.25}^{+1.07}$ & 64.62$_{-0.45}^{+0.51}$ & 159.02$_{-0.56}^{+0.42}$ &  55.82$_{-0.60}^{+0.58}$ &   7.26$_{-0.79}^{+0.73}$ &  &  &  &  \\
\hline
GO Tau & 1.3 mm & 54.59$_{-0.59}^{+0.69}$ & 170.64$_{- 4.83}^{+ 7.19}$ & 53.93$_{-0.60}^{+0.39}$ & 20.95$_{-0.57}^{+0.57}$ &  73.04$_{-0.64}^{+0.49}$ &   4.85$_{-1.85}^{+1.05}$ & 110.06$_{-1.03}^{+1.56}$ &   8.51$_{-1.90}^{+3.22}$ &  &  \\
GO Tau & 2.9 mm &  7.65$_{-0.13}^{+0.10}$ & 178.53$_{-12.41}^{+15.29}$ & 52.78$_{-0.59}^{+0.60}$ & 19.39$_{-0.87}^{+0.76}$ &  73.25$_{-0.82}^{+0.68}$ &   1.97$_{-0.57}^{+1.37}$ & 112.35$_{-1.70}^{+1.70}$ &   9.37$_{-2.36}^{+3.05}$ &  &  \\
\hline
DL Tau & 1.3 mm & 169.99$_{-0.72}^{+0.46}$ & 163.20$_{-1.18}^{+0.97}$ & 44.99$_{-0.14}^{+0.39}$ & 51.95$_{-0.33}^{+0.42}$ &  49.38$_{-0.97}^{+0.65}$ &   4.62$_{-0.99}^{+0.77}$ &  77.22$_{-0.37}^{+0.88}$ &   1.46$_{-0.67}^{+0.91}$ & 116.01$_{-0.57}^{+0.48}$ &   1.44$_{-0.14}^{+1.40}$ \\
DL Tau & 2.9 mm &  27.28$_{-0.08}^{+0.09}$ & 158.69$_{-1.35}^{+1.51}$ & 44.33$_{-0.16}^{+0.16}$ & 51.46$_{-0.21}^{+0.42}$ &  48.96$_{-0.54}^{+0.57}$ &   5.95$_{-0.85}^{+0.57}$ &  77.85$_{-0.18}^{+0.21}$ &   0.47$_{-0.08}^{+0.12}$ & 114.69$_{-0.66}^{+0.57}$ &   4.32$_{-1.51}^{+1.15}$ \\
\enddata
\tablecomments{(1)Target name. (2) Observed wavelength. (3) Disk mm flux. (4) Disk effective radius as 95\% of total disk flux encircled. (5) Disk inclination angle (0$\degr$ is face-on and 90$\degr$ is edge-on). (6) Disk position angle (east of north). (7)-(12) Radial location and width of dust rings as Gaussian sigma.
Adopted values are the peaks of the posterior distributions, with uncertainties representing the 68\% confidence interval and scaled by the square root of the reduced $\chi^2$ of the fit. Disk parameters for 1.3\,mm data are derived from new fitting with the same fitting setup as 2.9\,mm data. Ring widths here for DL Tau rings are smaller than what reported in \citet{Long2018}, because the prior on sigma was set to [0\farcs02, 0\farcs2]. Comparing to the fitting result in \citet{Long2018}, the derived new parameters fit better the 1.3\,mm DL Tau data with fewer residuals as seen in Figure~\ref{fig:uvmodel-app}.}
\end{deluxetable*}

\begin{figure}[!t]
\centering
    \includegraphics[width=\columnwidth]{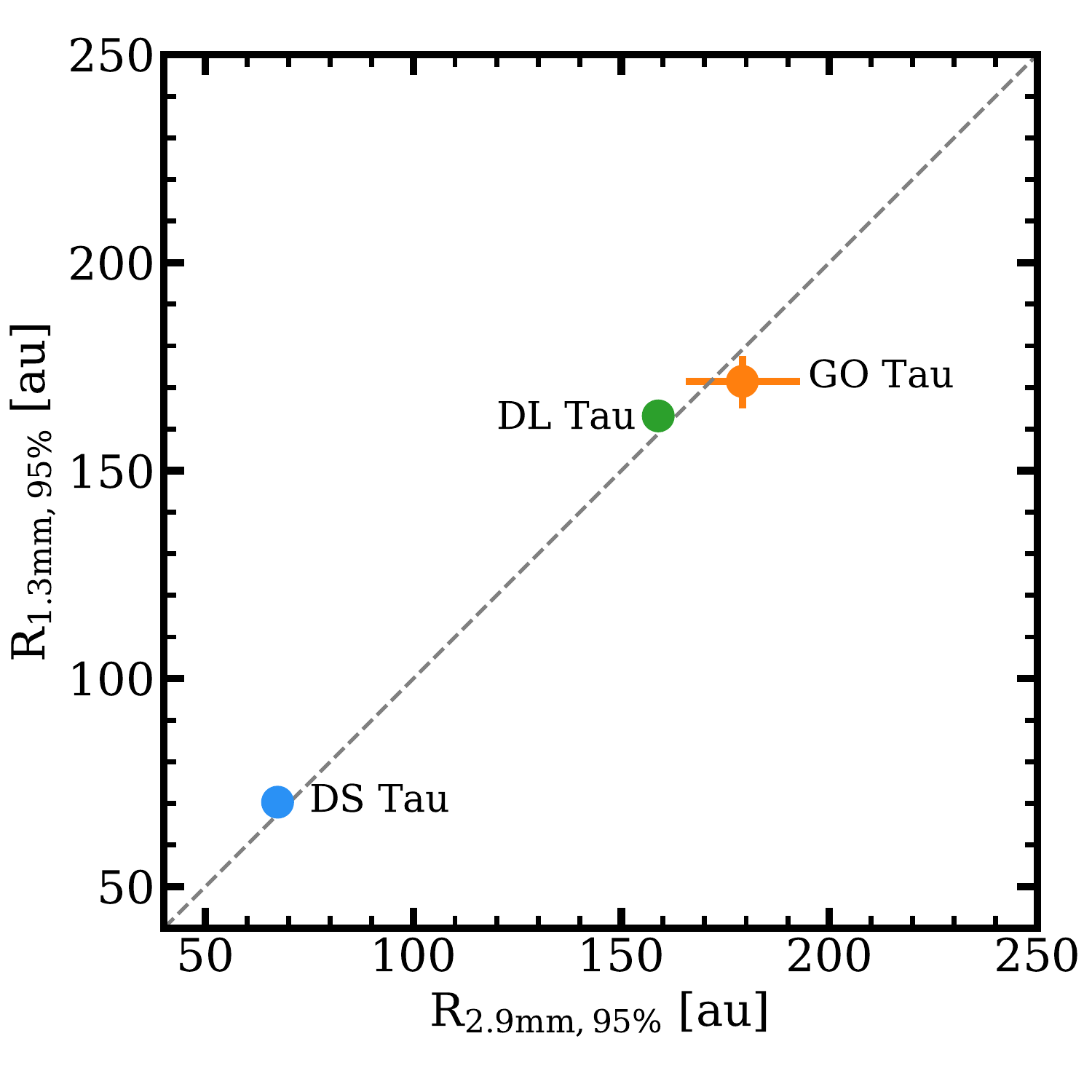} 
    \caption{Comparison of disk effective radii, defined as where 95\% of flux encircled from the intensity profile models, at 1.3 and 2.9\,mm. The errors for DS Tau and DL Tau are smaller than the symbol sizes. \label{fig:Rd_comp} }
\end{figure}

\subsubsection{Disk radius} \label{sec:radius}
Through visual inspection, dust emission is detected out to similar radial distances at 1.3 and 2.9\,mm for our individual disks. To quantify the disk radius, we adopt a generic definition of size - the location where a fixed fraction of the total disk flux is encircled, as introduced by \citet{Tripathi2017}. For our interest of the disk outer radius, we measure the effective disk size as 95\% of emission encircled in the adopted model intensity profile and estimate the uncertainties of R$_{\rm eff,95\%}$ as the 68\% confidence intervals from its posterior distribution. 

The comparisons of disk effective radius at 1.3 and 2.9\,mm are shown in Figure~\ref{fig:Rd_comp}. The measurements lie close to the 1:1 line in the plot, revealing consistent disk radii at both wavelengths. In DS Tau and DL Tau, R$_{\rm eff,95\%}$ at 1.3\,mm is slightly larger than that at 2.9\,mm by $\sim$3 and 5\,au, respectively, which are not statistically significant. The difference in DS Tau is mainly attributed to the subtle ring peak shift (by 0$\farcs$007, $\sim 1.2$\,au) and ring width change (by 0$\farcs$005, $\sim 0.8$\,au), in which the ring at 2.9\,mm is slightly narrower and located closer in (see more discussions about dust rings below). For DL Tau, one additional component is necessary to be included in the model to account for the faint fuzzy disk edge. For simplicity, we adopt a Gaussian ring to model the faint emission and this component takes up about 10\% of total disk flux and inevitably affects our disk radius measurement. This is similar to the case of GO Tau, which hosts a tenuous outer disk beyond the well-detected rings (R73 and R110) and requires an additional component in the fitting. Our fitting results in a larger disk radius (by $\sim 8$\,au, comparable to 1$\sigma$ uncertainty) at 2.9\,mm than at 1.3\,mm for GO Tau. This is because the Band\,3 fitting favors a very faint ring for the outer disk area (barely seen in the radial profile), wider than what we have obtained for the Band\,6 outer component, while our observations have very poor sensitivity at that radial distance. A better constraint on disk radius would be achieved with future higher sensitivity data. Overall, our observations demonstrate that disk radii are very similar at the two wavelengths.

The conclusion that the disks have similar radii at both 1.3 and 2.9\,mm holds as long as the adopted size metric includes the prominent dust rings that contribute to a significant fraction of the size metric encircled flux. Taking the single-ring system DS Tau as an example, the dust ring accounts for 50--60\% of the total dust emission, thus any metric larger than 50\% would result in similar disk radii. In the most extreme cases, transition disks, the majority of emission is confined to a specific radial range. The measured disk outer radii should therefore scale with the location of dust rings \citep[e.g.,][]{Andrews2018_Lmm}.

As slightly more compact emission in the inner disks is observed at the 2.9\,mm data, we also measure the spatial extent of the inner disk with the derived model intensity profiles. For this calculation, we only take the inner part of the intensity profile, which is cut at the first local minimum. The inner disk radius is then given as the radial location where 68\% of emission encircled. Our measurements show that $R_{\rm inner,2.9mm}$ is smaller than $R_{\rm inner,1.3mm}$ by 3-4\,au, with a typical 1$\sigma$ uncertainty of 0.6\,au. The values for $R_{\rm inner,2.9mm}$ and $R_{\rm inner,1.3mm}$ are 11.7 and 15.5\,au for DS Tau, 24.8 and 28.6\,au for GO Tau, and 24.8 and 27.8\,au for DL Tau.

\subsubsection{Dust rings}
The dust ring locations and widths (sigma of Gaussian ring) from visibility fitting at both Band\,3 and Band\,6 are summarized in Table~\ref{tab:uvmodel_prop}. The locations of individual rings are consistent at both wavelengths. By comparing the derived ring width with the beam size ($\sigma_{beam}=b_{fwhm}/2.355$, $\sim$7.5\,au at Taurus distance),
we find that all rings are spatially unresolved, except for the ring in DS Tau and the (very faint, thus less reliable) R110 ring in GO Tau, which are marginally resolved (ring width comparable to $\sim$1\,beam sigma). Our fitting results indicate narrower rings at longer wavelength for the R57 ring of DS Tau, the first ring (R73) of GO Tau, and the second ring (R77) of DL Tau (see more discussions of ring width fitting in the image profiles in Appendix~\ref{sec:gauss-fit}).

The difference of fitted ring width for the marginally resolved dust ring in DS Tau is, however, very subtle. Ring widths at both wavelengths are consistent within uncertainties, with a first hint for a narrower ring at 2.9\,mm than 1.3\,mm. This is based on the fitting result from the Gaussian profile plus Gaussian ring model. If we take the exponentially tapered power-law model for the inner disk, which matches better in the baseline range of 1000--1500 k$\lambda$ for both wavelengths (see Figure~\ref{fig:uvmodel-dstau-pl} in Appendix~\ref{sec:ds-PL}), the ring width stays the same as the Gaussian profile model at 1.3\,mm while it becomes narrower by more than 20\% at 2.9\,mm ($\sigma_{uv,\rm 2.9mm}= 5.4\pm0.8$\,au). The width difference is thus statistically significant. In addition, peak locations of the dust ring are better aligned at two wavelengths for the power-law model. Though the comparison demonstrates how the selected functional forms affect the fitting parameters, in this case both models prefer a slightly smaller dust ring at the longer wavelength.

As seen from the radial intensity profiles (see also Figure~\ref{fig:ring_image} in Appendix~\ref{sec:gauss-fit}), disk components in GO Tau and DL Tau are blended, indicative of the narrowness of the dust rings. In the fitting for both disks, one additional faint Gaussian ring is included to account for the tenuous outer disk edge. The fitting favors a broad component and overlaps with the interior dust rings, which could therefore affect (likely underestimate) the derived dust ring properties.
Meanwhile, a significant source of uncertainty in parametric fitting comes from the choice of functional forms, where systematic errors for disk properties of interest are hard to quantify. The ring width difference when comparing the Gaussian profile fit with the power-law fit for the DS Tau disk has already demonstrated how the choice of functions for the inner disk affects the connected dust rings in the outer disk.
Considering the complex dust morphology and the very likely presence of small-scale substructures in the inner disk, uncertainties in the dust ring properties could be largely underestimated in GO Tau and DL Tau. The derived values thus should be taken with caution. To quantify the real shape of these rings requires future higher angular resolution observations.

\citet{Dullemond2018} analyzed the high-contrast and well-separated rings in the DSHARP sample, which are spatially resolved with typical ring width of 3--7\,au for the sigma of Gaussian rings in the radial range of 40--120\,au. The unresolved nature of our rings are consistent with the narrow sizes found in the DSHARP rings. For the marginally resolved dust ring in DS Tau, the ring width is about twice the local pressure scale height ($h_p$), which is estimated as $\sqrt{\frac{k_{\rm B}T_{\rm d}r^3}{\mu m_{p}GM_*}}$, assuming the gas temperature is equal to dust temperature given in Section~\ref{sec:alpha}. As pointed out by \citet{Dullemond2018}, it is possible that strong turbulent mixing prevents the formation of even narrower dust rings, and/or drift-mixing equilibrium may not have been reached for dust grains responsible for our observed wavelengths.

\begin{figure*}[!t]
\centering
    \includegraphics[width=0.325\textwidth]{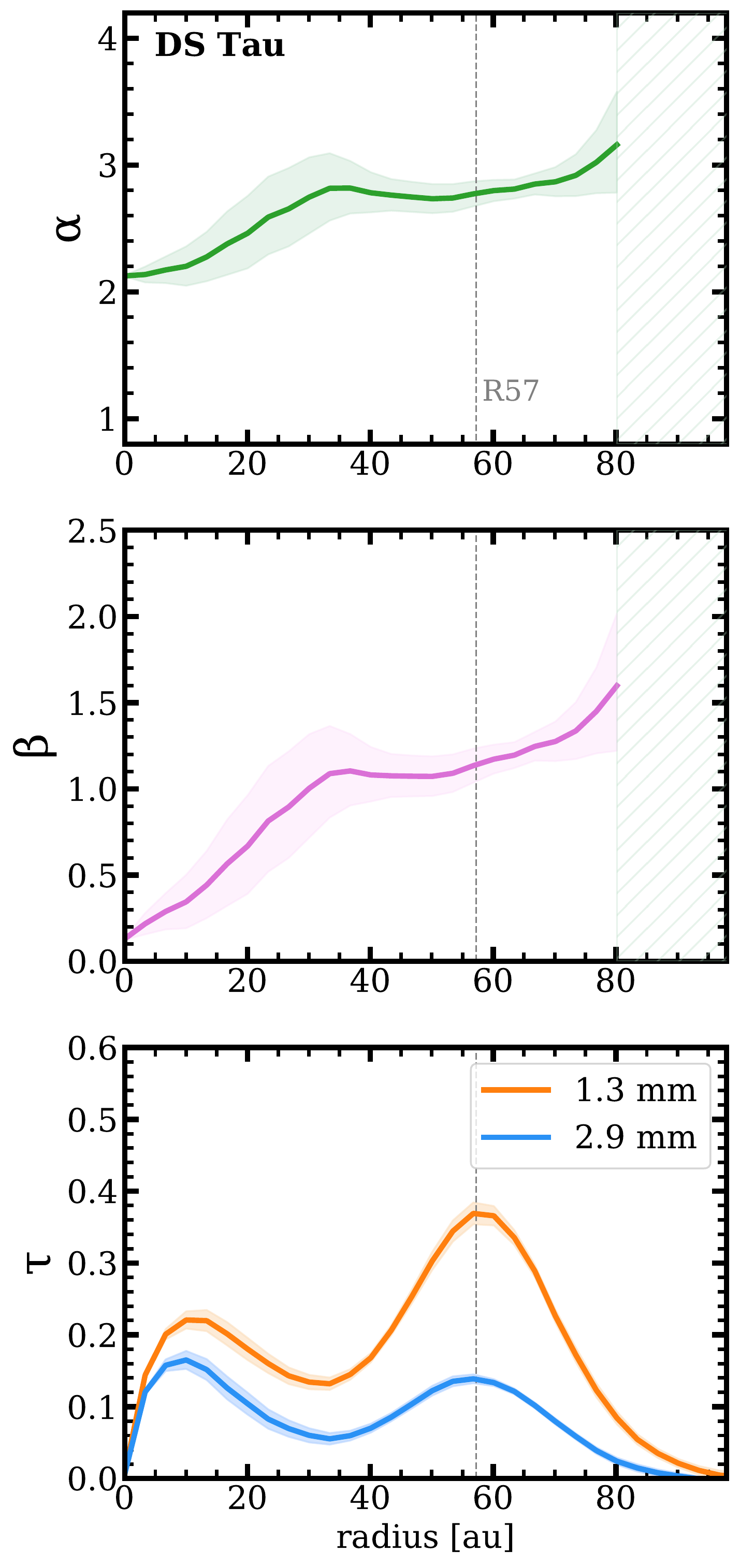}
    \includegraphics[width=0.325\textwidth]{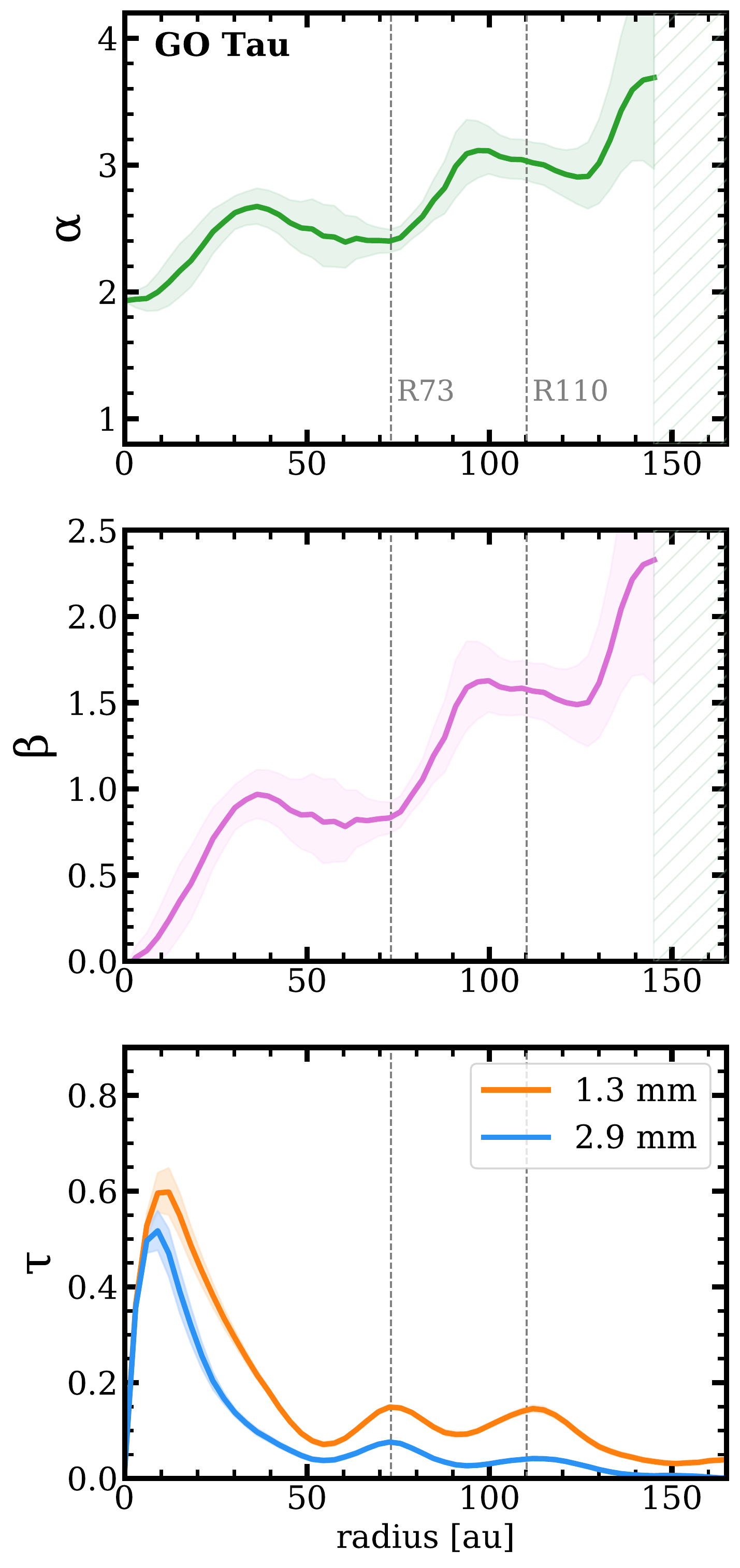}
    \includegraphics[width=0.325\textwidth]{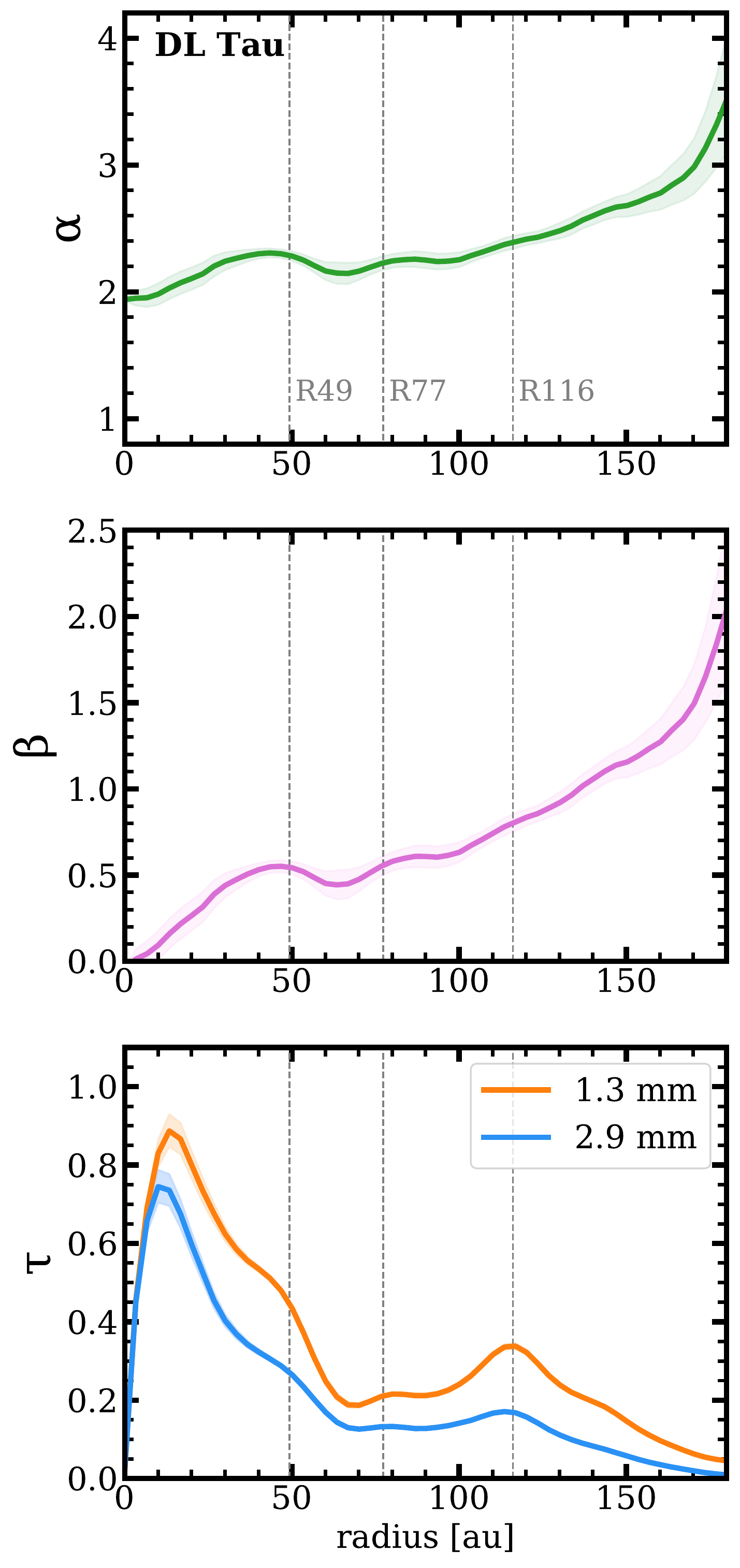} \\
    \caption{\textit{From top to bottom}: (1) Spectral index profiles, derived from radial profiles at two wavelengths. (2) Power-law index of the dust opacity dependence on frequency. (3) Radial profiles of continuum optical depth at two wavelengths. Calculations here use the full Planck expression. Light shaded regions show the 1$\sigma$ scatter divided by the square root of the number of beams spanning the full azimuthal angle at each radial bin. The typical beam size is 0$\farcs$1, corresponding to 15\,au. The absolute flux uncertainty is not accounted here. Regions with signal-to-noise ratio below unity are marked out. \label{fig:alpha} }
\end{figure*}

\subsection{Spectral index} \label{sec:alpha}
Based on the adopted model intensity profiles at two wavelengths, we derive the spectral index profile as 
\begin{equation}
    \alpha_{\rm mm}= log(I_{\nu_1}/I_{\nu_2})/log(\nu_1/\nu_2).
\end{equation}
As shown in the right panels of Figure~\ref{fig:uvmodel}, variations are seen across dust gaps and rings, where local minima in $\alpha_{\rm mm}$ are clearly observed around the dust ring of DS Tau and the R73 ring of GO Tau, the two high contrast rings in our sample. We also see local minima in other radii that can be easily produced due to the slight shifts in fitted gap and ring locations, leading to complex profile appearances and confusing the interpretation. Spectral index profiles estimated from the azimuthally averaged brightness profiles (see Figure~\ref{fig:alpha}) preserve the overall morphology as seen from the model profiles, but calculations from the image profiles largely damp the variation amplitudes and smooth out the sharp features.

In the analysis below, we adopt the spectral index profiles estimated from the images at the two wavelengths, as they show much cleaner patterns. Emission from the inner disks are likely optically thick as $\alpha_{\rm mm}$ is low ($\sim$2). We see an overall increasing trend in $\alpha_{\rm mm}$ with larger distance, reaching above 3 towards the outer disk. Similar to the model spectral index profiles, local minima (above 2) in $\alpha_{\rm mm}$ are observed around the peaks of high-contrast dust rings, though with large scatters ($\sim0.2$) around gap locations. Such variation is not seen across the rings of DL Tau. This is mainly due to the effect of observational resolution. Strong corresponding variations in spectral index have been reported in the dust gaps and rings in HL Tau and TW Hydra with 2--3\,au resolution \citep{alma2015, Tsukagoshi2016, Huang2018_CO}. By degrading the image of HL Tau to our resolution ($0\farcs1$, 15\,au), the abrupt changes across the gaps and rings would be largely suppressed, as also evident from the comparison of data and model spectral index profiles for our disks. The absolute value of $\alpha_{\rm mm}$ is therefore sensitive to resolution. Another source of uncertainty in the absolute value of $\alpha_{\rm mm}$ comes from the flux calibration uncertainty, in which a $\sim10$\% flux calibration uncertainty at each wavelength would introduce a systematic offset of $\sim$0.2 in $\alpha_{\rm mm}$. The disk-integrated $\alpha_{\rm mm}$ calculated from the total fluxes at the two wavelengths are 2.66, 2.58, and 2.40 for DS Tau, GO Tau, and DL Tau, respectively.

If dust emission is optically thin and dust opacity is dominated by absorption, the measured spectral index can be used to infer the dust grain properties (e.g., maximum grain size). 
We estimate optical depth of the dust emission using the expression of
\begin{equation}
    I_{\nu}(r)=B_{\nu}(T_d(r))(1-e^{-{\tau_{\nu}}}),
\end{equation}
where the full Planck function is adopted.
Since mm grains are largely settled to the disk midplane, the dust temperature is adopted as the disk midplane temperature using the simple irradiated flared disk assumption as 
\begin{equation}
    T_{\mathrm{d}}(r) = \left(\frac{\frac{1}{2}\varphi L_{*}}{4\pi r^2\sigma_{\mathrm{SB}}}\right)^{1/4}
\end{equation}
\citep[e.g.,][]{Chiang1997, DAlessio1998, Dullemond2001}. The flaring angle is taken to be $\varphi=0.02$, the same as the value used in the DSHARP analysis \citep{Huang2018_ring, Dullemond2018}, corresponding to T$_{d}\sim$\,15\,K and h/r = 0.06--0.08 at 50\,au for our sample. Since a larger flaring angle will lead to a warmer disk and lower value of optical depth, our choice of $\varphi$ results in a conservative temperature estimate and higher end of optical depth. As shown in Figure~\ref{fig:alpha}, the typical optical depth is about 0.1--0.2 for dust rings at 2.9\,mm and 0.2--0.4 for dust rings at 1.3\,mm. These values are broadly consistent with optical depth estimates in DSHARP rings.

The optical depth ($\tau_{\nu}$) is proportional to disk surface density and dust opacity ($\kappa_{\nu}$). At millimeter wavelength, $\kappa_{\nu}$ is often approximated as $\kappa_{\nu}=\kappa_0(\nu/\nu_0)^{\beta_{\rm mm}}$, in which the power-law index $\beta_{\rm mm}$ has strong dependence on the maximum grain size ($a_{\rm max}$) and the grain size distribution slope $q$, when dust opacity is absorption-dominated. In the optically thin case, $\beta_{\rm mm}$ can be directly related to $\alpha_{\rm mm}$ as $\beta_{\rm mm}=\alpha_{\rm mm}-log(B_{\nu_1}/B_{\nu_2})/log(\nu_1/\nu_2)$. 
The $\beta_{\rm mm}$ profiles are shown in Figure~\ref{fig:alpha}, presenting similar radial variations as $\alpha_{\rm mm}$ profiles. We find $\beta_{\rm mm}<0.5$ inside 20\,au and $\beta_{\rm mm}\sim$\,0.5--1.5 in our dust ring peaks (50--100\,au), which are lower than the expected $\beta_{\rm ISM}$
($\sim1.7$, \citealt{Li_Draine2001}) in interstellar medium where small $\mu$m-sized grains dominate. Low values of $\beta_{mm}$ have been reported from both disk-integrated and spatially-resolved measurements (e.g., \citealt{Beckwith1990, Ricci2010, Ricci2010_oph, Perez2012, Tazzari2016}) and are often interpreted as collisional growth of dust particles \citep[e.g.,][]{Liuyao2017}. 
The observed lower $\beta_{\rm mm}$ values in the dust rings indicate the presence of large particles, which could be the accumulation of drifting large dust grains from nearby regions or rapid grain growth in the higher density regions.
The overall value of $\beta_{\rm mm}$ (or $\alpha_{\rm mm}$) is lower in the brighter disk DL Tau, which may indicate faster grain growth in brighter disks, as also seen in the Lupus sample \citep{Ansdell2018}. 
We note that the uncertainties in $\alpha_{\rm mm}$ due to observational resolution and dust temperature also propagate to $\beta_{\rm mm}$, making the absolute spectral index less robust than the behavior of its radial variations, which reflects the radial change of dust opacity functions.

Dust opacity, as well as its spectral index, have complex dependence on dust grain sizes, chemical compositions and morphologies (shapes and internal structures) \citep[e.g.,][]{Miyake1993, Pollack1994, Draine2006, Kataoka2015_model}. The inference of maximum grain size ($a_{\rm max}$) from dust opacity index depends on the adopted dust model assumptions. A recent work by \citet{Birnstiel2018} discussed the effects of grain properties on dust opacity and provided a reference model for public use. At millimeter wavelength, when $a_{\rm max}>\lambda_{\rm obs}$, $\kappa_{\nu}$ decreases with $a_{\rm max}$, with a decline slope dependent on grain size distribution slope (see also e.g., \citealt{Ricci2010}). Based on the dust model of \citet{Birnstiel2018}, $\beta_{\rm mm}\sim$0.5--1.5 corresponds to $a_{\rm max}$ of mm or cm sizes for our disks.

The analysis above is based on an assumption that millimeter emission is dominated by absorption. Dust scattering is likely another major source of dust opacity, and it has recently received wide attention in interpreting observations at millimeter wavelength. 
The self-scattering of thermal dust emission was introduced to explain the orientation and degree of millimeter-wave polarization, with an interpretation that the maximum grain size is only $\sim$\,100$\mu$m \citep{Kataoka2016_HLTau,Lin2019}. 
The inclusion of dust scattering will make optically thick disk regions appear optically thin and lead to very low spectral index ($\alpha<2$) \citep{Zhu2019,LiuHY2019}. 
The inferences of the maximum grain size would therefore require a proper treatment of dust scattering \citep{Carrasco2019}.

\begin{figure*}
 \centering
 \tabcolsep=0.05cm 
   \begin{tabular}{cc}   
    	\includegraphics[width=9.0cm]{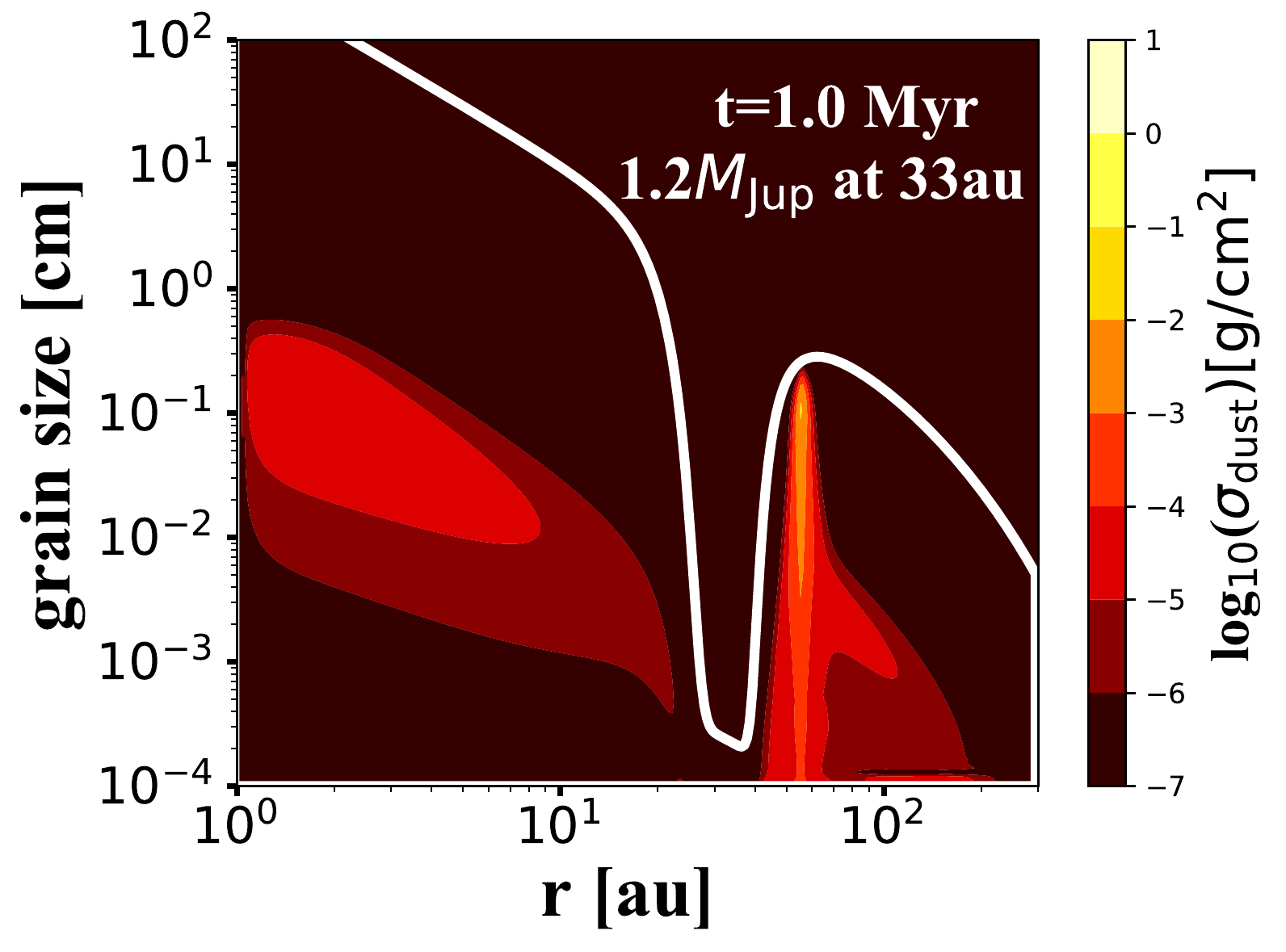}&
    	\includegraphics[width=8.5cm]{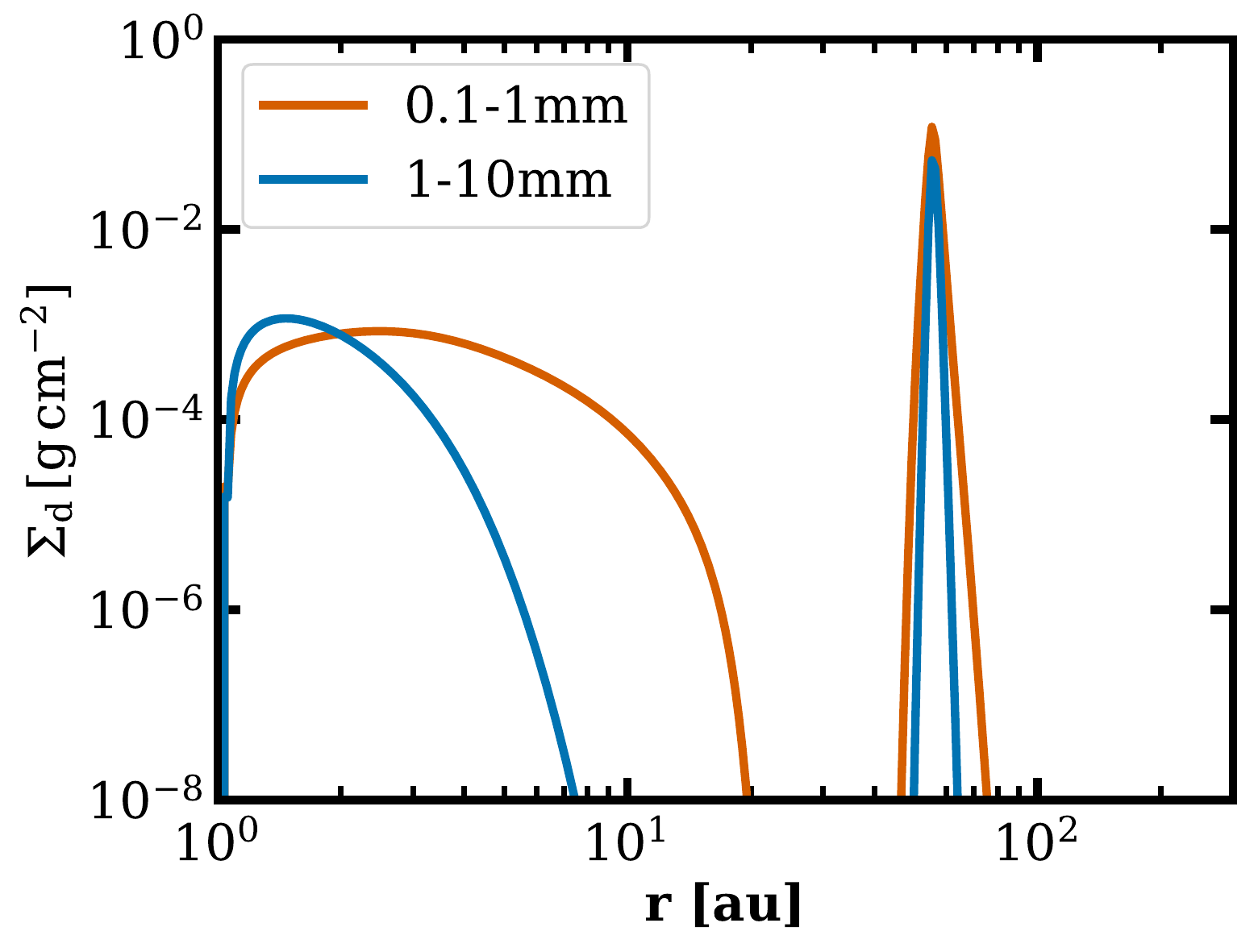}
    \end{tabular}
   \caption{Dust density distribution at 1\,Myr of evolution for the disk model with an embedded planet of $1.2\,M_{\rm{Jup}}$ mass at 33\,au. The left panel shows the total dust distribution where the white line corresponds to St=1 (representing the gas density profile). The right panel shows only the distribution of (sub-) millimeter-sized particles.  }
   \label{dust_distribution}
\end{figure*}

\section{Discussion} \label{sec:discussion}
In this section we present dust evolution models including grain growth, fragmentation, and radial drift for a disk with pressure bump introduced by an embedded planet. We investigate how millimeter-sized grains evolve in a timescale of 5\,Myr and compare to our observations.
We also explore the change in width of dust rings at different wavelengths and what constraints we can obtain on dust diffusion from ring width. Finally, dust disk sizes from observations over a wider range of wavelengths are discussed.

\subsection{Comparison with dust evolution models} \label{sec:dust-model}
One intriguing mechanism to produce the observed dust gaps and rings involves the interaction between planet(s) and the disk \citep[e.g.,][]{Lin1986,Pinilla2012_Planet,Zhu2012,Dipierro2015}. An embedded planet with sufficient mass creates a pressure bump outside the planet orbit, which traps large dust grains. A natural result of particle trapping would be keeping large grains of different sizes around the pressure bump, leading to similar dust disk at different (sub-)millimeter wavelengths (comparable to the size of large grains). We will show how dust disk radii evolve for grains with different sizes in a disk model with pressure bump introduced by an embedded planet and compare the results with a case of a smooth disk (no pressure bumps).

Since larger grains should drift more efficiently towards the local pressure maxima, a narrower ring at longer wavelength is expected \citep{Pinilla2015, Powell2019}. The dust concentration depends not only on the size of the particles (or their Stokes number), but also on the degree of dust diffusion \citep{Dullemond2018}. With dust evolution models, we could also provide some constraints on the dust diffusion with an assumed disk mass.

\begin{figure*}[!t]
 \centering
    \includegraphics[width=0.98\textwidth]{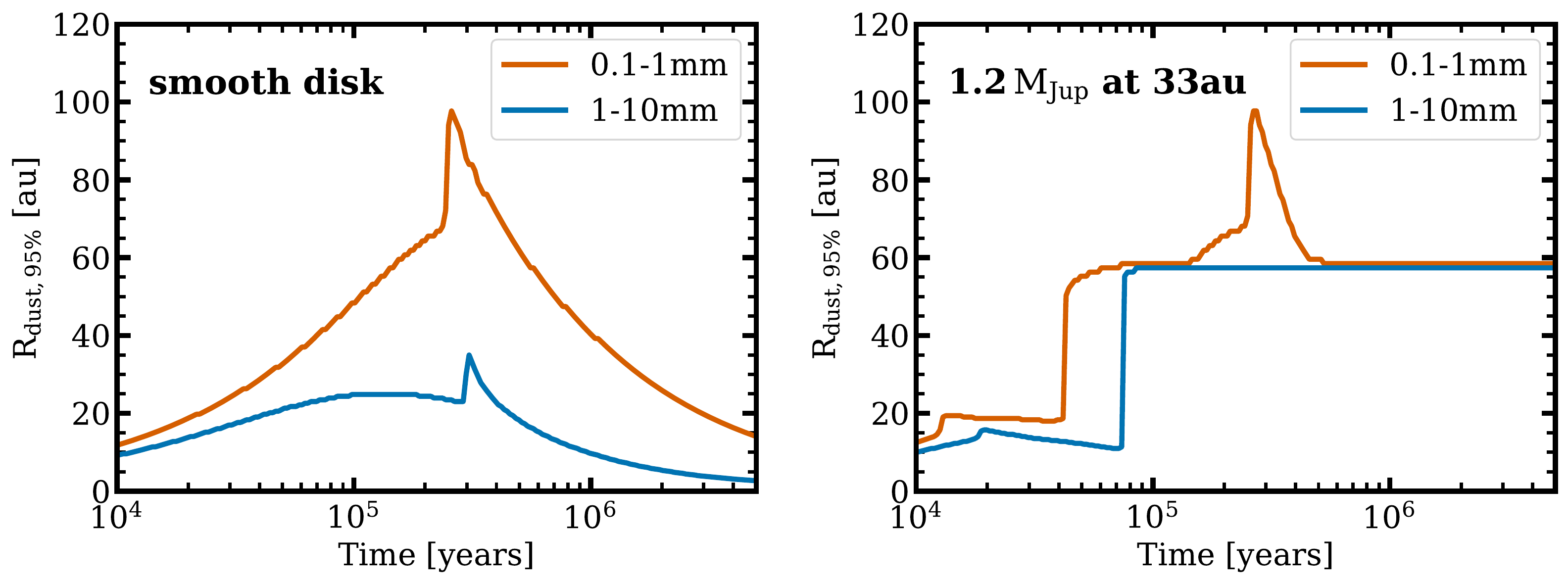} \\
\caption{Evolution of the radius that encloses 95\% of the mass of dust particles of 0.1-1\,mm and 1-10\,mm in size for a smooth disk model (left) and a disk model with an embedded planet at 33\,au (right). At around 0.2\,Myr, radial drift starts to dominate the grain evolution in the outer disk. }
   \label{Rdust_evolution}
\end{figure*} 

\subsubsection{Model setup}
Our model including one embedded planet is motivated by the single-ring system DS Tau. 
The stellar parameters are taken from Table~\ref{tab:source_prop}. For the disk properties, we assume a disk mass of 5.8\,$M_{\rm{Jup}}$ \footnote{As a test, we also performed simulations with a disk with twice the mass in the fiducial model and found similar results.}, 
from the dust disk mass obtained from the 1.3\,mm emission with a gas-to-dust ratio of 100 and assuming that around 20\% of the mass in dust have been lost due to radial drift by million-year timescales. 
The disk surface density distribution is assumed as an exponentially tapered power-law function, given by $\Sigma_{\rm{gas}}(r)=\Sigma_0\left(\frac{r}{R_c}\right)^{-\gamma} \exp\left[-\left(\frac{r}{Rc}\right)^{2-\gamma}\right]$, with $\gamma=1$ and $R_c=80$\,au. A gap is formed due to a planet located at 33\,au, the minimum of the gap as inferred from the visibility models of DS\,Tau. Assuming the ring peak (around 57\,au) traces the pressure maximum and a typical separation of $\sim$8--9$R_H$ between the planet location and the location of pressure maximum  \citep{Pinilla2012_Planet}, we obtain a star-to-planet mass ratio of $0.002$, which corresponds to a $\sim1.2\,M_{\rm{Jup}}$ around a 0.58\,$M_\odot$ star. 
\citet{Veronesi2020} has recently performed hydrodynamical simulations of the DS Tau ring and fitted both the 1.3mm and the 2.9 mm radial profiles using a slightly higher planet mass of $3.5\pm 1 M_{\rm Jup}$, corresponding to a separation between planet location and the ring peak of $\sim 7.3 R_H$, which is close to our assumed value here. On the other hand, \citet{Lodato2019} has used a more restrictive separation criterion and inferred an even higher planet mass for DS Tau. Note, however, that both papers assume a larger stellar mass. 
In our models, the temperature profile is assumed as above for optical depth calculations. We use the prescription from \citet{Crida2006} for an analytical shape of the gap carved by a planet and assume such gas density profiles to run the dust evolution. A correction for the gap depth is taken into account from \citet{Fung2014}.  We take an $\alpha$-viscosity parameter of $10^{-3}$ (independent of radius and time), which sets the dust diffusion, settling, and turbulent velocities accordingly in the dust evolution models.

All grains are initially micron-sized particles that grow, fragment or erode due to mutual collisions. Fragmentation and erosion occurs when particles reach a fragmentation velocity that we set to 10\,m\,s$^{-1}$ \citep{Birnstiel2010}. The grid for particle size has 180 cells logarithmic spaced from 1\,$\mu$m to 2\,m, and the radial grid (300 cells) is also logarithmic spaced from 1 to 300\,au. The dust density distribution for our model is shown in the left panel of Figure~\ref{dust_distribution}, in which the solid white line corresponds to the Stokes number equal unity at the midplane. The Stokes number quantifies the aerodynamical drag of particles, defined in the midplane as St$=a\rho_s\pi/2\Sigma_g$, where $\rho_s$ is the intrinsic volume density of the grains, set to 1.2\,g\,cm$^{-3}$, $a$ is the grain size, and $\Sigma_g$ is the gas surface density. Thus, St=1 line represents the gas distribution in the disk.

\subsubsection{Dust disk radius}
Figure~\ref{dust_distribution} (right panel) shows the radial distribution of grains from 0.1--1\,mm and 1--10\,mm, which are roughly the grain sizes that dominate the emission at 1.3 and 2.9\,mm, respectively. Dust grains for the two populations span similar radial extents, and peak around the pressure maximum ($\sim$57\,au). The disk size would be similar at the wavelengths that are sensitive to these two populations of grain sizes. 
We have also run a smooth disk model without the planet, in which the radial distributions of larger grains are more compact, similar to what is seen in the inner disk of the planet disk model.

\begin{figure*}
 \centering
    \includegraphics[width=0.98\textwidth]{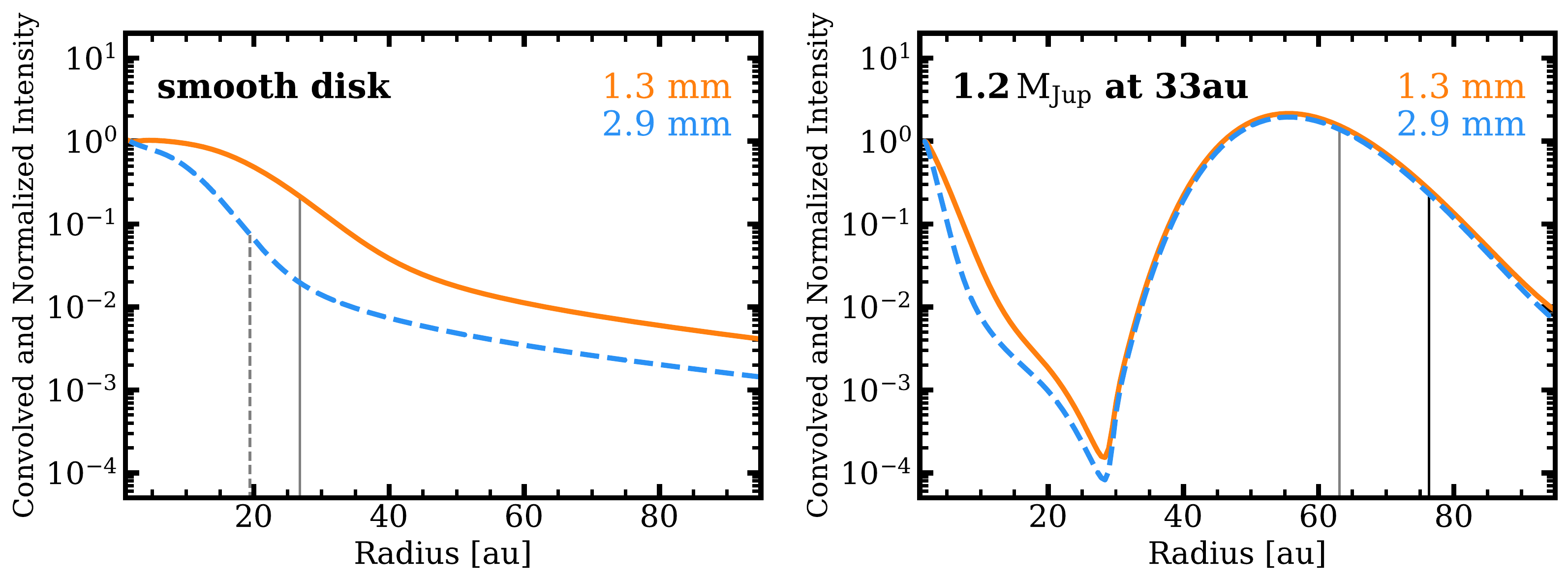} \\
   \caption{Intensity profile at 1.3\,mm and 2.9\,mm from the dust evolution models at 1\,Myr. The profiles are convolved with a Gaussian profile whose width is 0$\farcs$1 as our observations. Left panel shows the case of a smooth disk and the right panel shows the case with a planet. The vertical grey lines corresponds to the radius that encircles 68\%, while the vertical black line corresponds to the 95\% (solid line for 1.3\,mm and dashed line for 2.9\,mm). In the case of the smooth disk the 95\% radii lies outside the radial range and for the case of the planet the dashed and solid lines overlap because the disk radii does not change with wavelength. }
   \label{intensity_smooth_planet}
\end{figure*}

The evolution of the disk radius that encloses 95\% of the mass of dust particles of 0.1--1\,mm and 1--10\,mm in sizes is shown in Figure~\ref{Rdust_evolution}. The case without a gap and otherwise identical initial conditions is shown for comparison. The evolution of dust extension is determined by the competition of grain growth and (regulated) radial drift. The extension of the population of 0.1--1\,mm-sized particles starts with increasing up to around 100\,au till 0.3\,Myr as particles at larger radial distances take longer time to grow. 
Soon after radial drift dominates the grain evolution in the outer disk,  the disk extension of this dust population decreases with time in both models, while disk extension in the planet disk model converges to around 60\,au (just outside the pressure maximum) after 0.6\,Myr since they all get trapped at the pressure maximum. The disk radius for 1--10\,mm particles also starts with increasing from collisional growth, but radial drift dominates at a closer radius for this dust population that can not exist beyond 40\,au in our smooth model. 
Around the pressure maximum after 0.1\,Myr, the particles of 1--10\,mm sizes form locally and are trapped there. The disk radius for 1--10\,mm particles therefore stays around the pressure maximum. The maximum grain size outside 60\,au is limited by radial drift and grains of these sizes cannot form there. 
At the typical disk ages in Taurus (1--5\,Myr), we would expect that the disk radius does not change significantly between 1.3\,mm and 2.9\,mm if there is a trap in the outer disk. In contrast, in a smooth disk, the disk radius should become smaller at longer wavelength. This applies to our measurements of the inner disk radius (though within the beam).

We create intensity profiles at 1.3 and 2.9\,mm assuming the dust density distribution from the models with the dust opacity given by \citet{Ricci2010} (similar to DSHARP opacity from \citealt{Birnstiel2018}). Figure~\ref{intensity_smooth_planet} shows the intensity profiles at 1.3 and 2.9\,mm from the dust evolution models at 1\,Myr, which are convolved with a Gaussian profile whose width is 0\farcs1 as our observations. Both R$_{\rm eff,95\%}$ and R$_{\rm eff,68\%}$ are consistent at the two wavelengths when a pressure bump is present. In this case, the dust disk size is therefore regulated by the location of the local pressure maximum, different from a smooth disk where the disk size is drift-dominated \citep{Rosotti2019_RL}. This result is very important to understand the disk size distribution and evolution observed in nearby star-forming regions as well as the outer edge of our Solar System \citep{Hendler2020arXiv}.
Although our model is based on planet-disk interaction, other mechanisms capable of creating pressure gradient in the gas disk \citep{Johansen2009,Flock2015} could also reach the same conclusion. In addition, how the amplitude of gas density perturbation (strength of dust trapping) affects the dust dynamics in detail is still an open question, as low mass planets can form dust gaps without altering significantly the gas structure \citep{dipierro17}. 
These disks with extended diffuse emission in the outer disk region (e.g., GO Tau in our sample) may apply to the scenario with weak gas density perturbation.

\subsubsection{Dust ring width}
In the disk model with an embedded planet, the ring-like structure becomes slightly narrower for larger particles. From the unconvolved model intensity profiles, we measure the width of the dust ring as the standard deviation for a Gaussian distribution. We therefore determine the ring width to be only 1.09\,au at 2.9\,mm and 1.29\,au at 1.3\,mm. Observationally distinguishing this difference would be very challenging, although this difference strongly depends on the parameters of the model, in particular on the Stokes number (which depends on grain size, gas surface density, and intrinsic volume density of the particles), the shape of the pressure bump, and the $\alpha$ parameter that controls the dust diffusion,  settling,  and turbulent velocities. All of these parameters are still unknown from current observations.
Comparing the convolved intensity profiles at 1.3 and 2.9\,mm (Figure~\ref{intensity_smooth_planet}), the dust ring profiles basically overlap at both wavelengths and any difference in ring width is washed out.
In our visibility fitting, we find very similar dust ring width for DS Tau, with a weak hint for a slightly narrower ring at longer wavelength.  
The subtle width difference obtained for the ring of DS Tau is likely due to the full spatial information employed in the fitting, which would have an effective beam size smaller than 0$\farcs$1, and/or the exact shape of the uv-plane coverage between observations at two bands.

The width of our ring from the convolved profile of dust evolution models ($\sim8$\,au) broadly agrees with the width of the ring in DS\,Tau from the uv-fitting models, providing hints that the assumed $\alpha$ of $10^{-3}$ may be a good assumption for the disk viscosity and turbulence. However, the width of the ring remains marginally resolved from our observations, and any interpretation has to be taken with caution.
In our models, the width of the emission at the two wavelengths depends on the St/$\alpha$ ratio as demonstrated in \citet{Dullemond2018}. Increasing $\alpha$ in our simulations will result in a disk with higher viscosity, dust diffusion and turbulence, which will affect the capacity of planet of a given mass to open a gap and the efficiency of trapping, probably ending in a disk without visible structures as pointed out by \citet{deJuanOvelar2016}. Decreasing the value of $\alpha$ would imply that the concentration of the grains at pressure maximum is more effective for the same Stokes number, making the ring structure narrower. Because trapping is efficient for particles with St$\gtrsim\alpha$ \citep{Birnstiel2013}, if $\alpha$ is very low, grains of different size will be efficiently trapped in the pressure bump.

\subsection{Dust disk radii by different tracers}
Similar disk dust extensions at close wavelengths (between 0.9 or 1.3 and 2.9\,mm) are seen in previous studies \citep{alma2015, Tsukagoshi2016, Macias2019}. A recent work by \citet{Powell2019} measured disk radial extents in the dust continuum at a wider range of wavelengths (0.8--10\,mm) for a set of seven disks that most of them are known to have substructures. 
In the five disks with available measurements, they found disk radii at 0.9 or 1.3 and 2.9\,mm are mostly consistent within 1$\sigma$ uncertainties. Among them, only CY Tau has no reported dust substructures so far, while the large disk size of CY Tau likely indicates the presence of substructures \citep{Long2019}. FT Tau shows a notable size difference of 96\,au at 1.3\,mm and 60\,au at 2.6\,mm, each with an uncertainty of 10--20\,au \citep{Powell2019}. Recent high resolution observation at 1.3\,mm for FT Tau \citep{Long2018} has revealed a dust ring at 32\,au and a full disk within 60\,au. The discrepancy might indicate the existence of an extended diffuse outer disk beyond the detected dust ring, while this ring at 32\,au may also work as a dust trap given the disk sizes at 8--10\,mm wavelengths of 30--36\,au.

\citet{Powell2019} also finds large differences of disk radii when comparing measurements from short wavelengths ($\sim$1\,mm) and long wavelengths ($\gtrsim$7\,mm). A more compact disk at longer wavelength are usually taken as the observational evidence of radial drift and grain growth. According to our models, the dust disk radii should not change between wavelengths if there are pressure bumps at the outer disk to efficiently trap the dust particles responsible for the observed emission. The disk effective radius at 7\,mm from our model with a pressure bump is highly consistent with the radii measured for 1.3 and 2.9\,mm emission.
A possible explanation for a different dust disk radii at different wavelengths  in the presence of pressure bumps  could be that the pressure bumps formed late and after large grains have already drifted inwards \citep[e.g.][]{pinilla2015_hd100}.  Alternatively, the long wavelength observations reported in \citet{Powell2019} may not be sensitive enough to detect the cold large grains located in the outermost pressure bump with low surface brightness. This scenario corresponds to what \citet{Tripathi2018} proposed to explain the observed dust disk size--frequency relation in UZ Tau E disk that continuum emission is mostly optically thick in the inner disk and becomes optically thin in the outer disk. Optically thin outer regions need to be observed at high sensitivity to properly detect the outer radius at long wavelengths, otherwise similar dust disk sizes would be expected if dust emission is optically thick overall.

\section{Summary} \label{sec:sum}
This paper presents ALMA continuum observations at 2.9\,mm for three disks with detected dust rings at 1.3\,mm. The new ALMA observations are conducted at comparable angular resolution ($\sim0\farcs1$) to the previous 1.3\,mm measurements. 
The main goal is to explore the grain properties and dynamics by comparing the dust emission morphology at two wavelengths. Our key results are summarized as follows: 

\begin{enumerate}
\item Dust rings are detected at both wavelengths at corresponding locations for individual disks. 
For all three disks, the inner disks (with radius of 20--40\,au) are slightly more compact (by 3-4\,au) at the longer wavelength, an observational feature predicted by radial drift and grain growth models.

\item Disk models with pressure bumps predict narrower rings at longer wavelengths, but this subtle difference of ring width ($\sim$0.2\,au) at 1.3 and 2.9\,mm from our model for DS Tau is impossible to detect with $0\farcs1$ resolution. However, the difference of ring width at the two wavelengths depends on a number of parameters in the models such as the grain size, gas surface density, shape of the pressure bump, and viscosity; all of them being unknown by current observations.  The dust ring in DS Tau is marginally resolved, and shows a tentatively narrower ring at 2.9\,mm than at 1.3\,mm based on visibility fitting with sub-beam resolution.

\item GO Tau and DL Tau are multi-ring systems in which dust rings are largely unresolved. The derived ring width suffers from large uncertainties due to both the complex emission morphology and the choice of number and forms of model functions, making the comparison of ring width unfeasible.

\item Dust emission at both wavelengths have similar outer radii in our sample of disks, measured from the adopted model intensity profiles with parametric fitting in the uv-plane. This result is consistent with dust evolution models for disks with a pressure bump (e.g., caused by an embedded planet), which sets the disk outer radius. In a disk with smooth surface density distribution (lack of pressure bump), the disk would become more compact at longer wavelengths due to radial drift.

\item Radial profiles of spectral index ($\alpha_{\rm mm}$) show a general increasing trend towards outer disks and local variations across dust gaps and rings.  Local minima in $\alpha_{\rm mm}$ profile correspond to peaks of high-contrast rings (the dust ring in DS Tau and R1 ring in GO Tau). 

\item The inner disks with $\alpha_{\rm mm}$ reaching 2 are likely optically thick. Dust rings have typical optical depth of 0.1--0.2 at 2.9\,mm and 0.2--0.4 at 1.3\,mm. If optically thin emission is a reasonable assumption, grain growth should occur faster at high density rings and have already produced mm-sized particles. 

\end{enumerate}


\paragraph{Acknowledgments}
We thank the referee for comments that greatly improved this paper. We thank Karin Oberg, Richard Teague and Jane Huang for helpful discussions.
F.L. acknowledges support from the Smithsonian Institution as a Submillimeter Array (SMA) Fellow. F.L. and G.J.H. acknowledge general grant 11473005 awarded by the National Science Foundation of China. P.P. acknowledges support provided by the Alexander von Humboldt Foundation in the framework of the Sofja Kovalevskaja Award endowed by the Federal Ministry of Education and Research. D.H. is supported by the European Union A-ERC Grant 291141 CHEMPLAN, NWO and by a KNAW professor prize awarded to E. F. van Dishoeck. D.H. acknowledges support from the EACOA Fellowship from the East Asian Core Observatories Association. D.J. is supported by NRC Canada and by an NSERC Discovery Grant. 
ER acknowledges financial support from the European Research Council (ERC) under the European Union's Horizon 2020 research and innovation programme (grant agreement No 681601).

This paper makes use of the following ALMA data: 2016.1.01164.S and 2018.1.00614.S. ALMA is a partnership of ESO (representing its member states), NSF (USA), and NINS (Japan), together with NRC (Canada), MOST and ASIAA (Taiwan), and KASI (Republic of Korea), in cooperation with the Republic of Chile. The Joint ALMA Observatory is operated by ESO, AUI/NRAO, and NAOJ.

\facilities{ALMA}
 

\software{analysisUtils~(\url{https://casaguides.nrao.edu/index.php/Analysis_Utilities}), AstroPy~\citep{Astropy2013}, CASA~\citep{McMullin2007}, emcee~\citep{ForemanMackey2013}, Galario~\citep{Tazzari2018},matplotlib~\citep{matplotlib2007} Scipy~(\url{http://www.scipy.org)}  }

\bibliography{ms}{}
\bibliographystyle{aasjournal}

\appendix

\section{Visibility fitting results} \label{sec:uvfit-plots}
\begin{figure*}[!h]
\centering
    \includegraphics[width=0.48\textwidth, trim=0 0 0 50, clip]{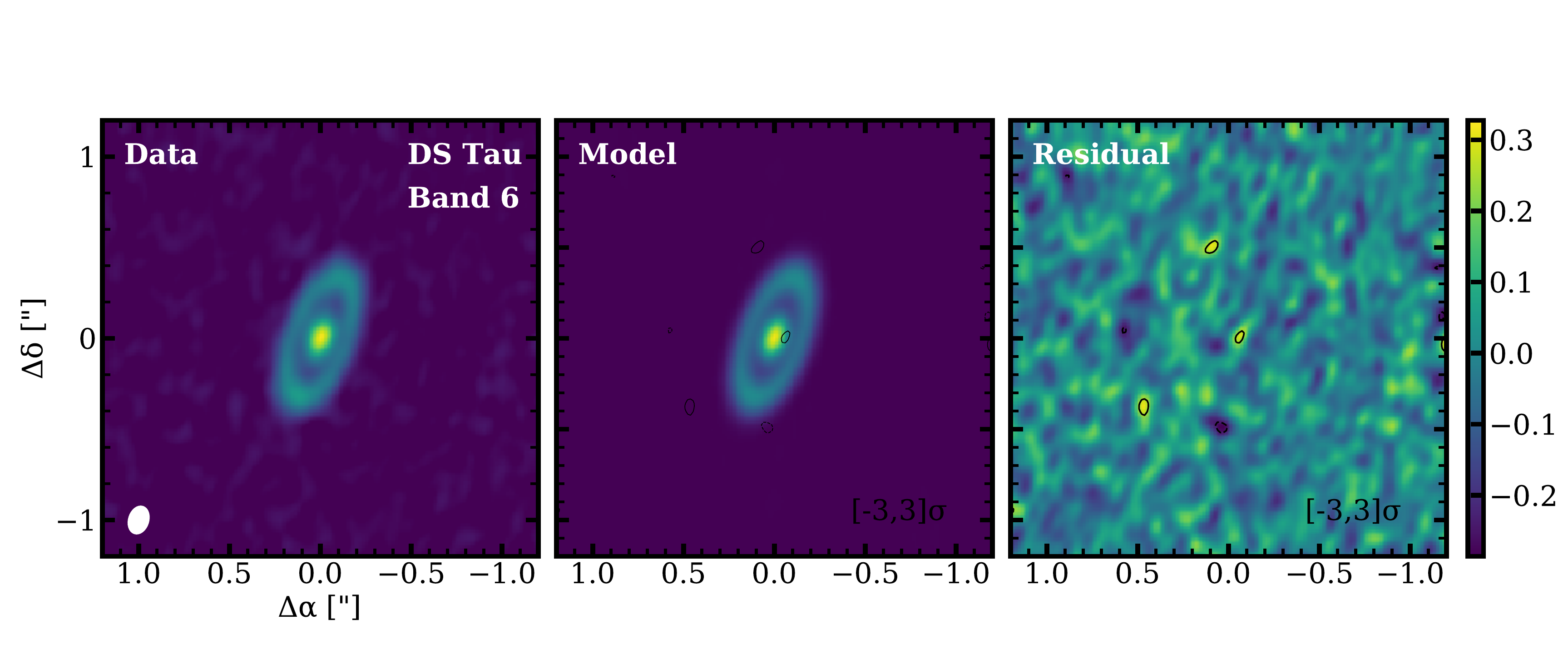}  
    \includegraphics[width=0.48\textwidth, trim=0 0 0 50, clip]{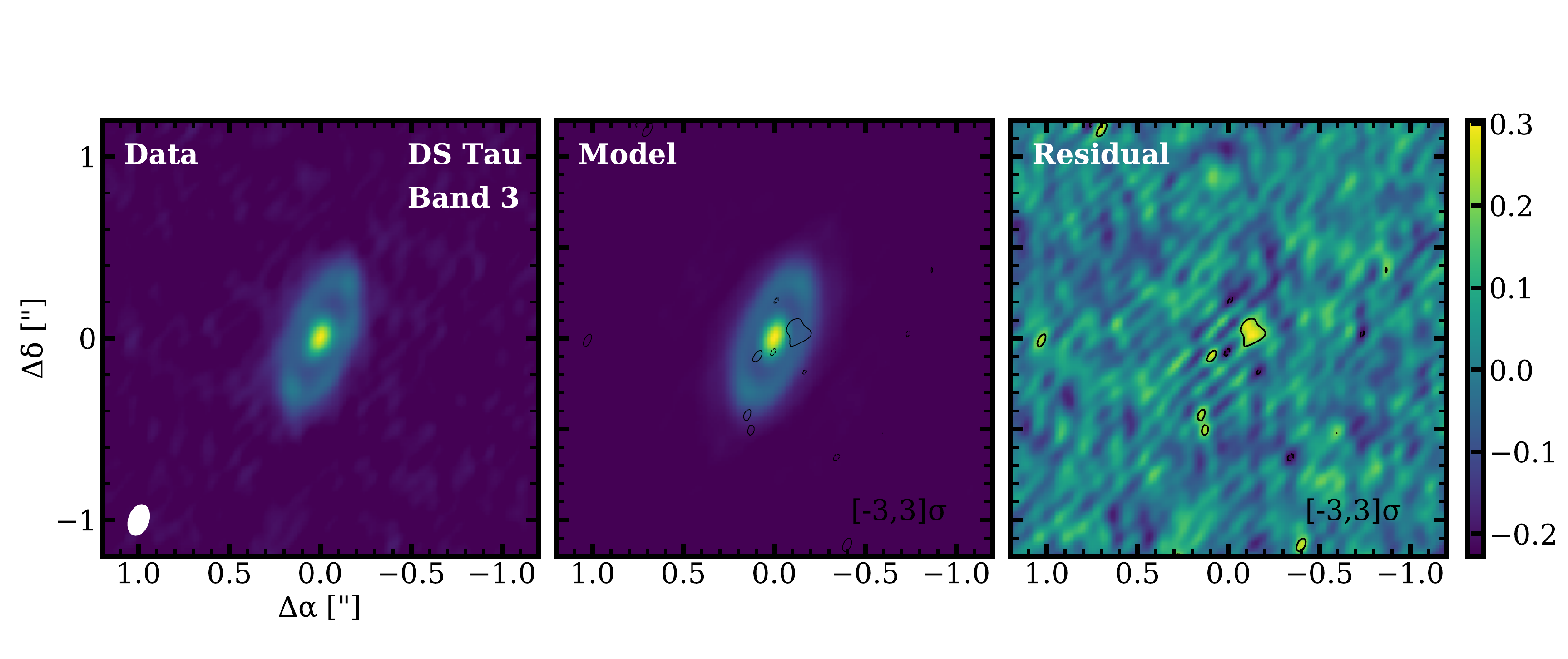} \\
    \includegraphics[width=0.48\textwidth, trim=0 0 0 50, clip]{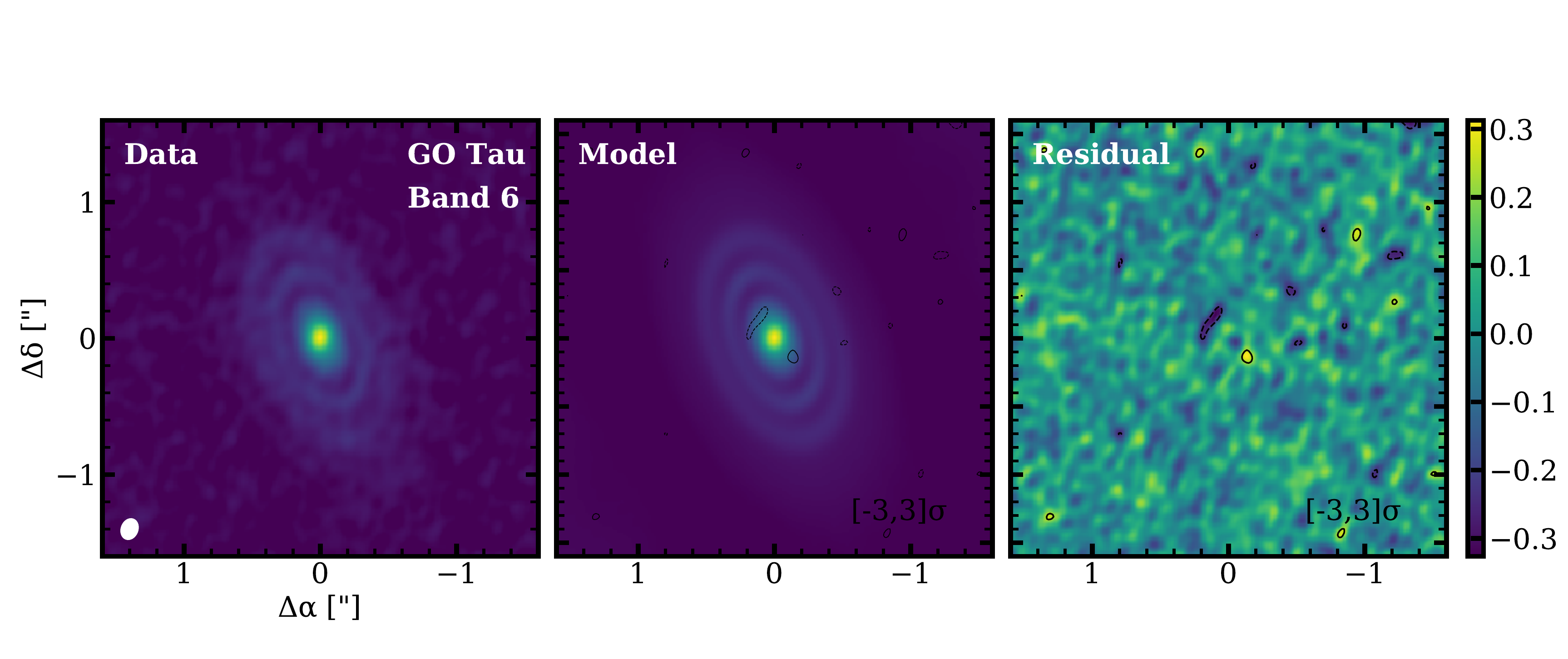}  
    \includegraphics[width=0.48\textwidth, trim=0 0 0 50, clip]{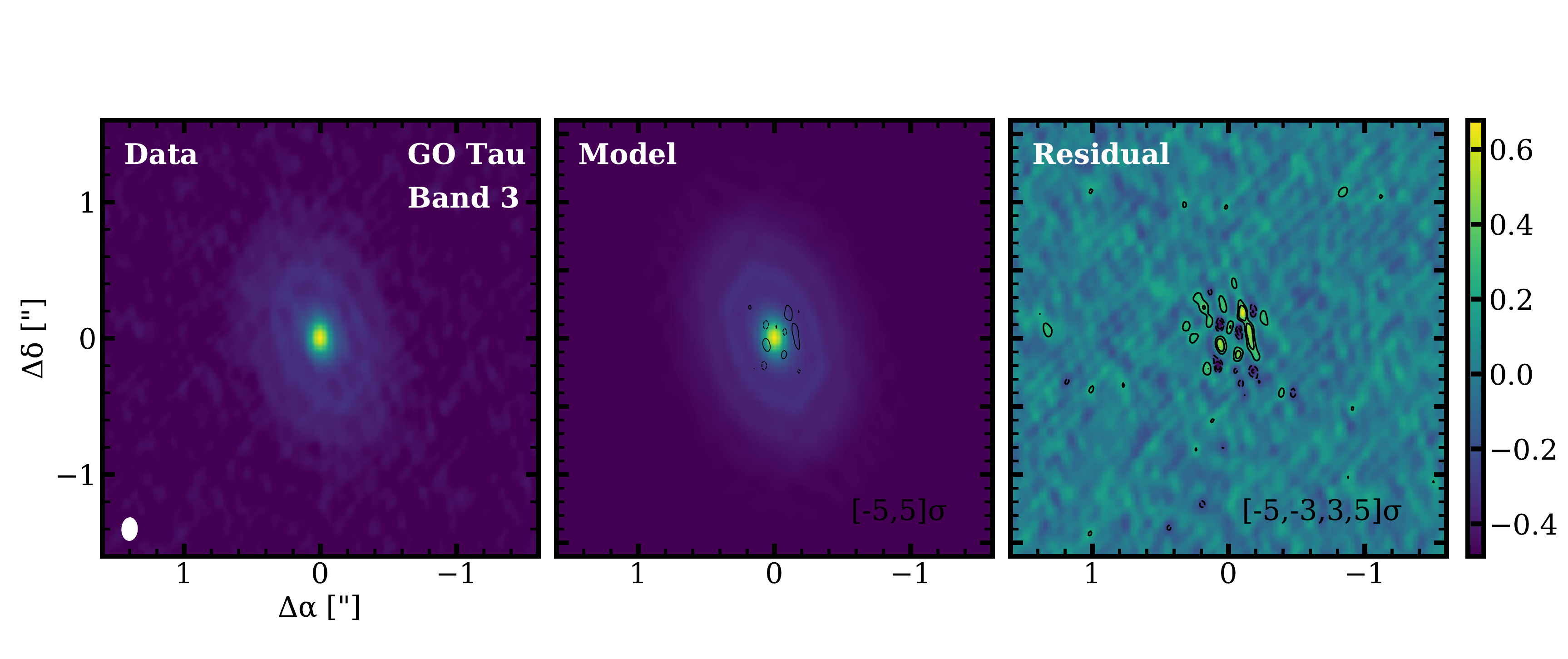} \\
    \includegraphics[width=0.48\textwidth, trim=0 0 0 50, clip]{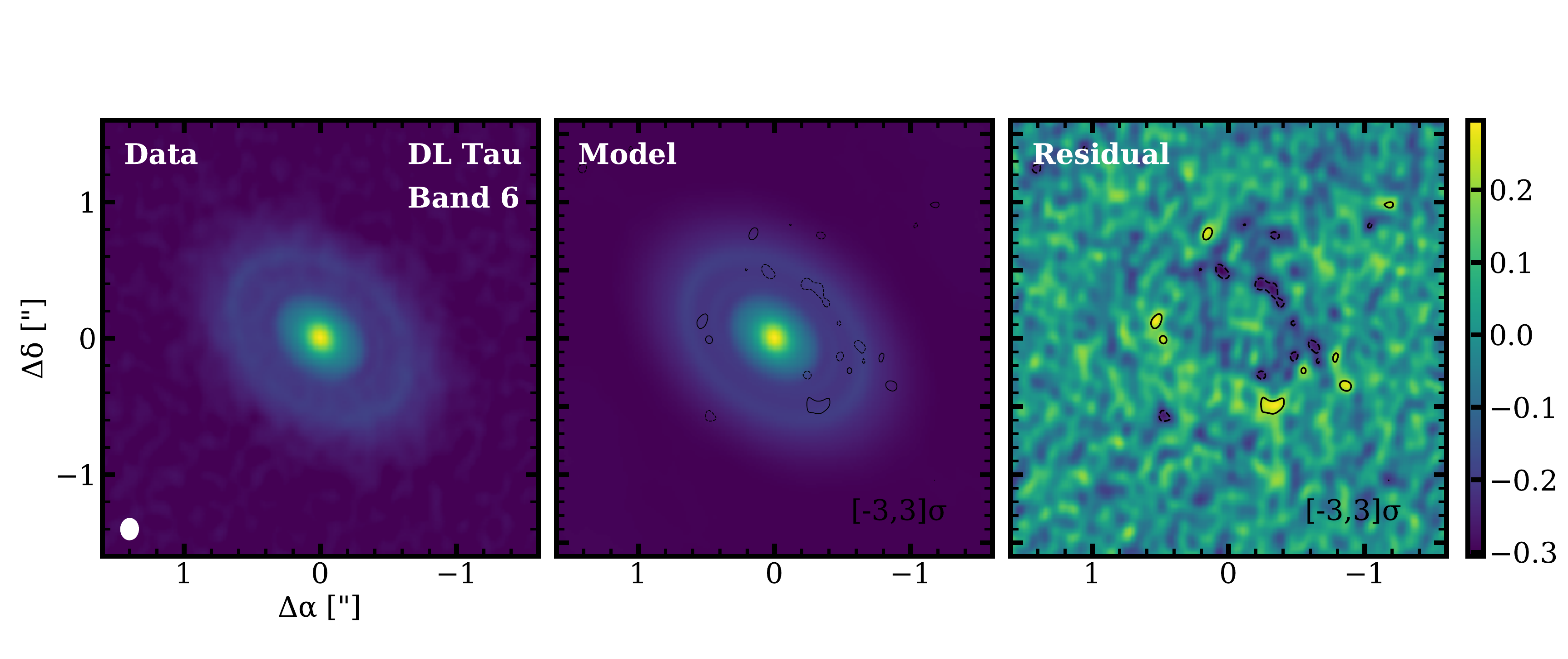} 
    \includegraphics[width=0.48\textwidth, trim=0 0 0 50, clip]{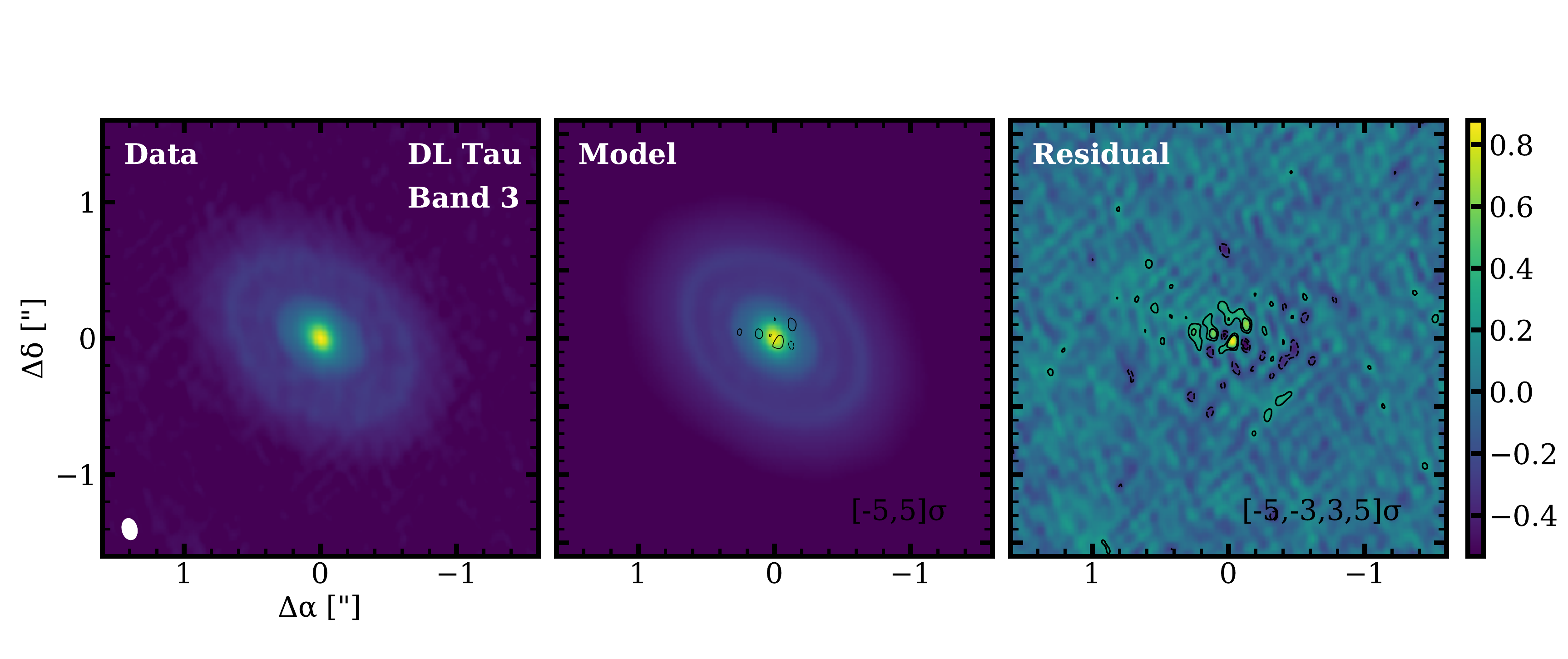} \\
    \caption{Comparison of the data and adopted model for three disks at both wavelengths. Colorbar is in unit of brightness temperature for the residual map. \label{fig:uvmodel-app} }
\end{figure*}

Figure~\ref{fig:uvmodel-app} shows the synthesized images for the data, the adopted model, and the residual. For each disk, same intensity functional forms are adopted. While the maximum residual in the image is about 3$\sigma$ for Band\,6 data, higher ($\sim5\sigma$) residuals are seen for Band\,3 data. Model parameters for individual disks are summarized in Tables~\ref{tab:dstau_uvmodel_prop}, \ref{tab:gotau_uvmodel_prop}, \ref{tab:dltau_uvmodel_prop}. 

\begin{figure*}[!h]
\centering
    \includegraphics[width=0.95\textwidth, trim=0 0 0 0, clip]{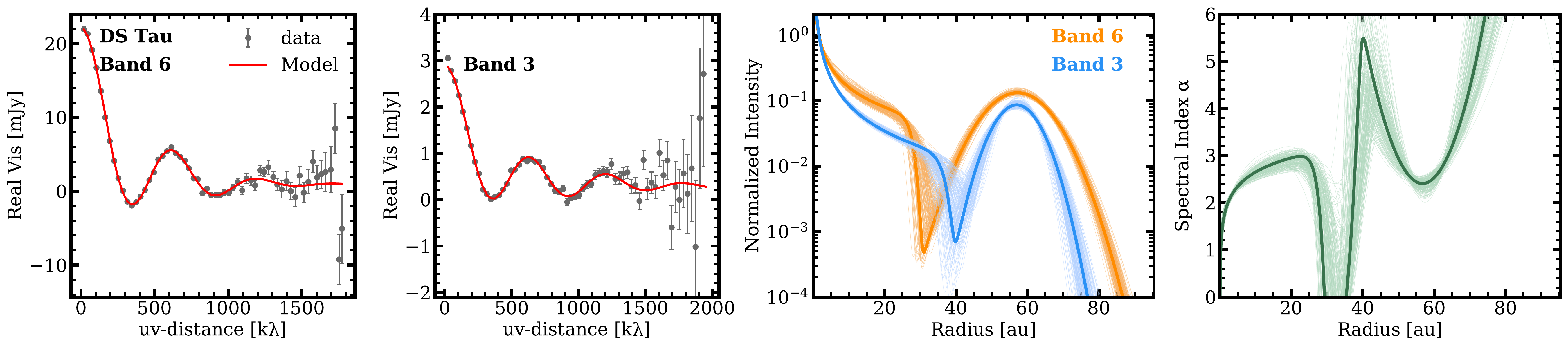}  \\
    \includegraphics[width=0.48\textwidth, trim=0 0 0 50, clip]{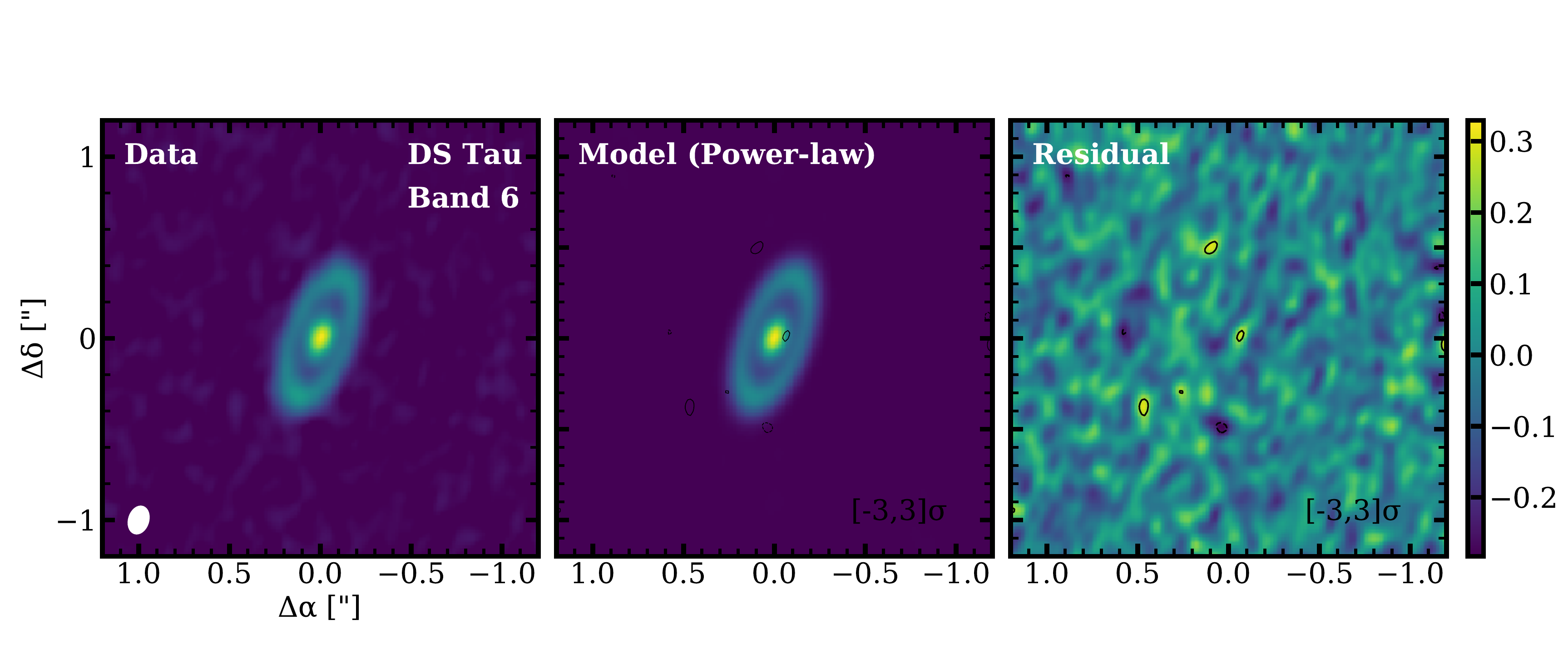}  
    \includegraphics[width=0.48\textwidth, trim=0 0 0 50, clip]{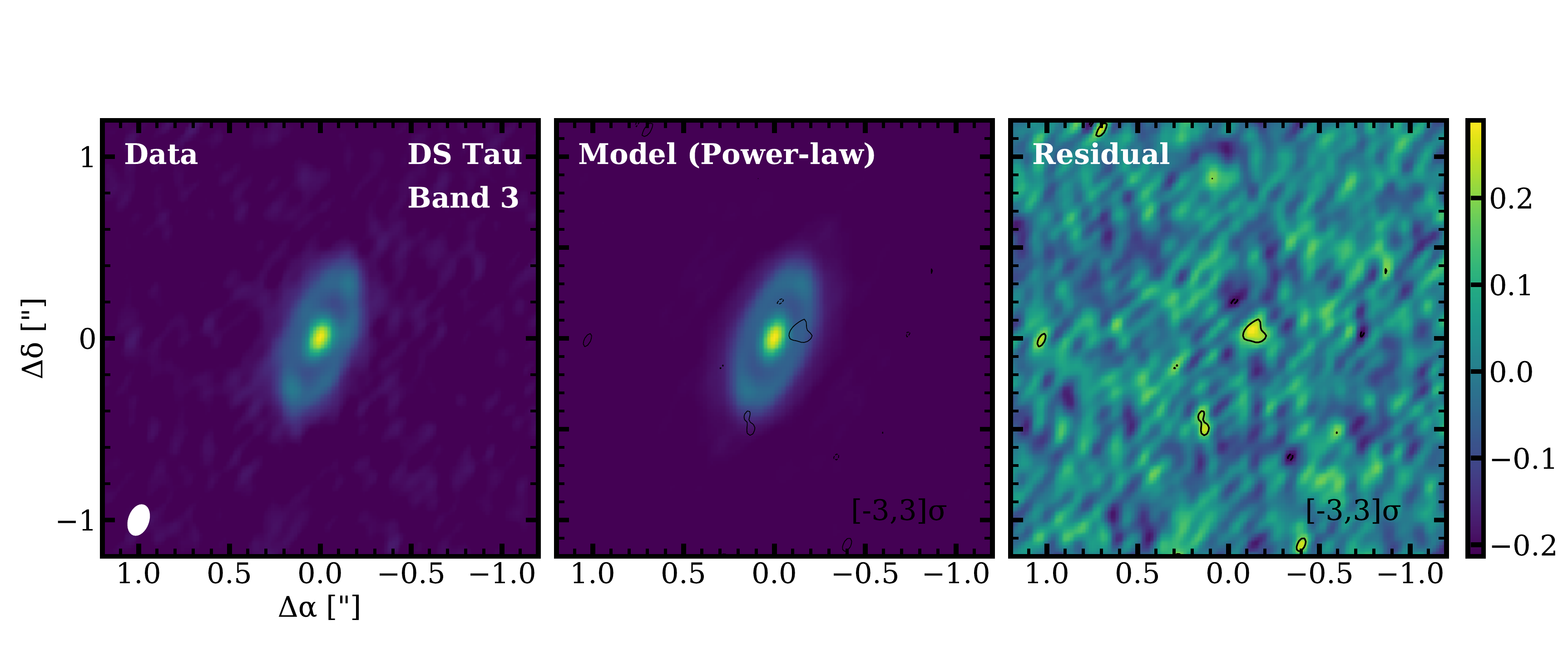} \\
    \caption{Comparison of the data and adopted power-law model for DS Tau at both wavelengths.  \label{fig:uvmodel-dstau-pl} }
\end{figure*}

\begin{deluxetable*}{lccccccccc}[!h]
\tabletypesize{\scriptsize}
\tablecaption{Model results for DS Tau \label{tab:dstau_uvmodel_prop}}
\tablewidth{0pt}
\tablehead{
\colhead{Band} & \colhead{F$_{0}$} & \colhead{$\sigma_{0}$ (R$_{c}$)} & \colhead{$\gamma_1$} & \colhead{$\gamma_2$} & \colhead{F$_{1}$} & \colhead{$\sigma_1$} & \colhead{R$_{1}$} & \colhead{$\delta$RA} & \colhead{$\delta$Dec} \\
} 
\startdata
\multicolumn{10}{c}{Gaussian Profile + Gaussian Ring} \\ 
\hline
1.3mm & 10.293$_{-0.033}^{+0.023}$ & 0.065$_{-0.003}^{+0.003}$ & & & 9.777$_{-0.018}^{+0.020}$ & 0.051$_{-0.003}^{+0.002}$ & 0.359$_{-0.002}^{+0.003}$ &  -0.135 & 0.218 \\ 
2.9mm &  9.747$_{-0.039}^{+0.043}$ & 0.049$_{-0.003}^{+0.003}$ & & & 8.905$_{-0.035}^{+0.044}$ & 0.046$_{-0.005}^{+0.005}$ & 0.351$_{-0.004}^{+0.004}$ &  -0.178 & 0.158 \\ 
\hline
\multicolumn{10}{c}{Tapered Power-law + Gaussian Ring} \\ 
\hline
1.3mm & 9.424$_{-0.140}^{+0.122}$ & 0.176$_{-0.010}^{+0.029}$ &  1.003$_{-0.121}^{+0.086}$ & 13.384$_{-6.065}^{+4.558}$ & 9.789$_{-0.017}^{+0.025}$ & 0.049$_{-0.004}^{+0.003}$ & 0.360$_{-0.002}^{+0.003}$ & -0.135 & 0.219 \\ 
2.9mm & 8.233$_{-0.125}^{+0.114}$ & 0.232$_{-0.023}^{+0.039}$ &  1.338$_{-0.054}^{+0.035}$ & 13.691$_{-5.344}^{+4.334}$ & 8.990$_{-0.040}^{+0.078}$ & 0.034$_{-0.006}^{+0.004}$ & 0.358$_{-0.004}^{+0.004}$ & -0.179 & 0.157 \\ 
\enddata
\tablecomments{Model parameters for both Gaussian profile and Power-law profile are listed. The definition for each parameter can be found in Section~\ref{sec:morphology}. Uncertainties for phase center offsets are about 0.001 and not shown here.}
\end{deluxetable*}

\begin{deluxetable*}{lcccccc}
\tabletypesize{\scriptsize}
\tablecaption{Model results for GO Tau \label{tab:gotau_uvmodel_prop}}
\tablewidth{0pt}
\tablehead{
\colhead{Band(Ring Index)} & \colhead{F$_{0}$} & \colhead{$\sigma_{0}$} & \colhead{$\gamma_1$} & \colhead{$\gamma_2$} & \colhead{$\Delta$RA} & \colhead{$\Delta$Dec} \\
} 
\startdata
1.3mm & 9.456$_{-0.013}^{+0.021}$ & 0.317$_{-0.011}^{+0.006}$ & 1.057$_{-0.014}^{+0.014}$ & 6.929$_{-0.784}^{+1.042}$ &  -0.168 & -0.405 \\
R1 & 9.403$_{-0.062}^{+0.096}$ & 0.034$_{-0.013}^{+0.007}$ & 0.507$_{-0.004}^{+0.003}$ & & & \\
R2 & 8.970$_{-0.089}^{+0.045}$ & 0.059$_{-0.013}^{+0.022}$ & 0.764$_{-0.007}^{+0.011}$ & & & \\
R3 & 8.237$_{-0.050}^{+0.319}$ & 0.330$_{-0.203}^{+0.051}$ & 1.011$_{-0.503}^{+0.058}$ & & & \\
\hline
2.9mm & 9.011$_{-0.174}^{+0.103}$ & 0.220$_{-0.027}^{+0.035}$ & 1.070$_{-0.042}^{+0.064}$ & 1.290$_{-0.146}^{+0.336}$ &-0.175 & -0.430 \\
R1 & 8.804$_{-0.177}^{+0.195}$ & 0.014$_{-0.004}^{+0.009}$ & 0.509$_{-0.006}^{+0.005}$ & & & \\
R2 & 7.958$_{-0.070}^{+0.079}$ & 0.065$_{-0.016}^{+0.021}$ & 0.780$_{-0.012}^{+0.012}$ & & & \\
R3 & 7.119$_{-0.116}^{+0.411}$ & 0.532$_{-0.220}^{+0.060}$ & 0.703$_{-0.333}^{+0.452}$ & & & \\
\enddata
\tablecomments{In each segment, the first row lists parameters for the inner disk with a tapered power-law profile and phase center offsets. The subsequent rows list the parameters for each Gaussian ring in order of amplitude, sigma, and location. The definition and unit for each parameter can be found in Section~\ref{sec:morphology}. Uncertainties for phase center offsets are about 0.001 and not shown here.}
\end{deluxetable*}

\begin{deluxetable*}{lcccccc}
\tabletypesize{\scriptsize}
\tablecaption{Model results for DL Tau \label{tab:dltau_uvmodel_prop}}
\tablewidth{0pt}
\tablehead{
\colhead{Band(Ring Index)} & \colhead{F$_{0}$} & \colhead{$\sigma_{0}$} & \colhead{$\gamma_1$} & \colhead{$\gamma_2$} & \colhead{$\Delta$RA} & \colhead{$\Delta$Dec} \\
} 
\startdata
1.3mm & 9.980$_{-0.022}^{+0.027}$ & 0.276$_{-0.021}^{+0.009}$ & 0.737$_{-0.022}^{+0.013}$ & 5.155$_{-0.441}^{+0.653}$ & 0.236 & -0.059 \\ 
R1 & 9.893$_{-0.066}^{+0.042}$ & 0.029$_{-0.006}^{+0.005}$ & 0.311$_{-0.006}^{+0.004}$ & & &\\
R2 & 9.727$_{-0.190}^{+0.219}$ & 0.009$_{-0.004}^{+0.006}$ & 0.486$_{-0.002}^{+0.005}$ & & &\\
R3 & 9.663$_{-0.281}^{+0.070}$ & 0.009$_{-0.001}^{+0.008}$ & 0.730$_{-0.003}^{+0.003}$ & & &\\
R4 & 9.318$_{-0.007}^{+0.010}$ & 0.227$_{-0.003}^{+0.010}$ & 0.674$_{-0.013}^{+0.005}$ & & & \\ 
\hline
2.9mm & 9.397$_{-0.041}^{+0.033}$ & 0.230$_{-0.008}^{+0.011}$ & 2.434$_{-0.182}^{+0.362}$ & 0.710$_{-0.017}^{+0.016}$ & 0.257 & -0.054 \\ 
R1 & 8.971$_{-0.024}^{+0.037}$ & 0.037$_{-0.005}^{+0.004}$ & 0.308$_{-0.003}^{+0.004}$ & & &\\
R2 & 9.458$_{-0.114}^{+0.070}$ & 0.003$_{-0.001}^{+0.001}$ & 0.490$_{-0.001}^{+0.001}$ & & &\\
R3 & 8.585$_{-0.049}^{+0.144}$ & 0.027$_{-0.010}^{+0.007}$ & 0.721$_{-0.004}^{+0.004}$ & & &\\
R4 & 8.450$_{-0.020}^{+0.023}$ & 0.257$_{-0.011}^{+0.015}$ & 0.607$_{-0.022}^{+0.018}$ & & & \\ 
\enddata
\tablecomments{In each segment, the first row lists parameters for the inner disk with a tapered power-law profile and phase center offsets. The subsequent rows list the parameters for each Gaussian ring in order of amplitude, sigma, and location. The definition and unit for each parameter can be found in Section~\ref{sec:morphology}. Uncertainties for phase center offsets are about 0.001 and not shown here.}
\end{deluxetable*}

\section{The Tapered Power-law fitting for DS Tau} \label{sec:ds-PL}
The mismatch of data and model in the visibility profile around 1200\,k$\lambda$ (Figure~\ref{fig:uvmodel}) for DS Tau implies sharper features in the disk than what can be described by the simple Gaussian profile and Gaussian ring model. We have performed testing fits using a tapered power-law function for the inner disk. The fitting results are shown in Figure~\ref{fig:uvmodel-dstau-pl}, with model parameters listed in Table~\ref{tab:dstau_uvmodel_prop}. The discrepancy at long baselines are largely resolved with this power-law model. A Nuker profile would also fit this long baselines oscillation equally well and not shown here.  The effects of the choice of different function forms on disk parameters are discussed in Section~\ref{sec:morphology}.

\section{Simple Gaussian fit to dust rings} \label{sec:gauss-fit}
\begin{figure*}[!h]
\centering
    \includegraphics[width=0.8\textwidth]{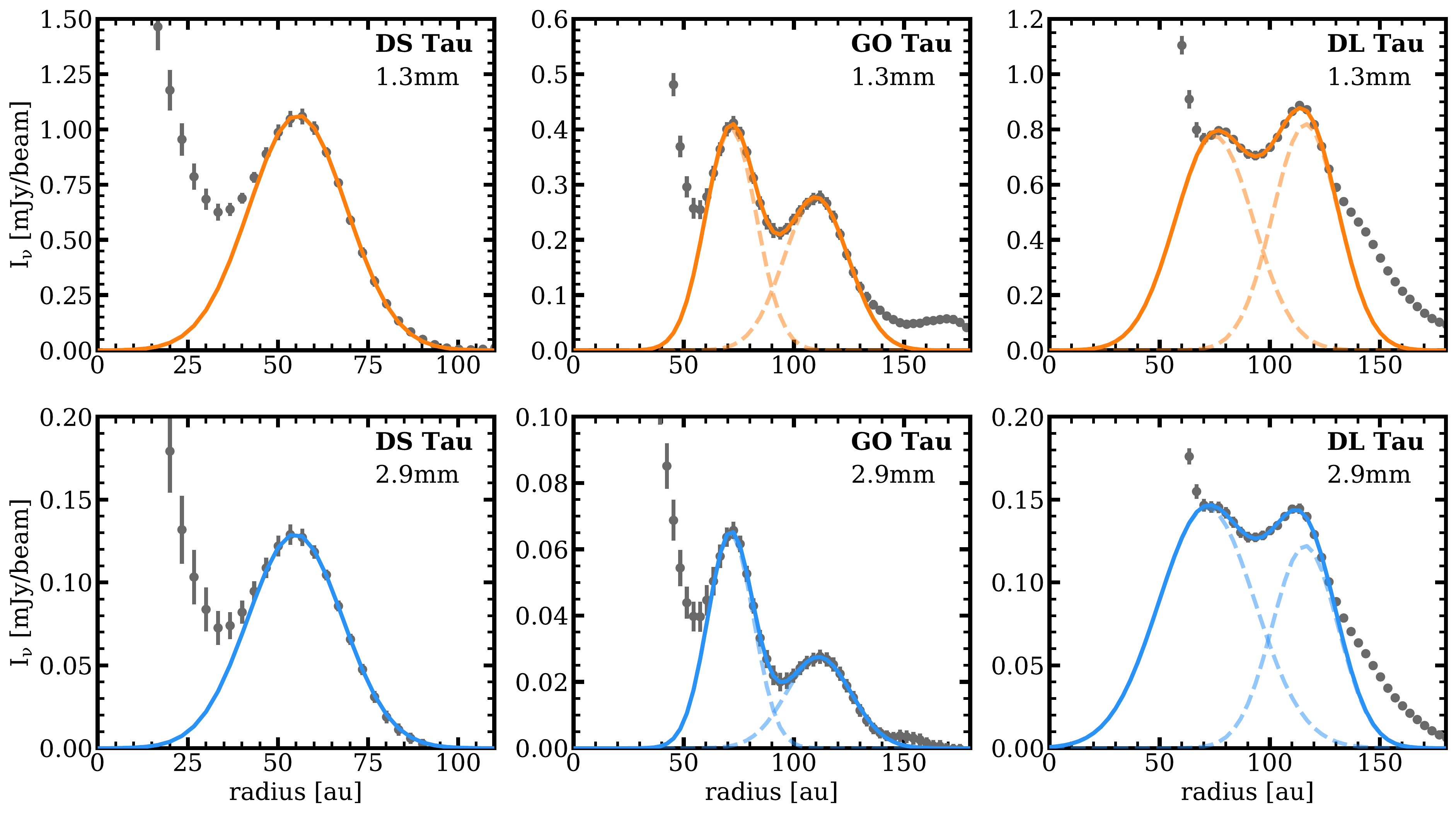} 
    \caption{Simple Gaussian fits to dust rings in the radial profiles. Data points are shown in grey dots and best-fit curves are overplotted in color lines. For GO Tau and DL Tau, the combined fit results are shown in solid lines and individual fit results are shown in dashed lines. \label{fig:ring_image} }
\end{figure*}

We provide simple Gaussian profile fit to individual dust rings using the azithmual averaged radial profiles to demonstrate the complexity of dust emission morphology. As disk components are not well separated, we only perform the fit within a limited radial range around the ring peak for each dust ring (see Table~\ref{tab:im_ring_prop}). The fit is performed with \textsc{$curve\_fit$} in \textsc{scipy} package. The underlying dust ring width is then derived by deconvolution from the Gaussian beam, specifically $\sigma_{de}=\sqrt{\sigma_{im}^2 - \sigma_{beam}^2}$. The fitting results are summarized in Table~\ref{tab:im_ring_prop}. Summing up the contributions from individual Gaussian components in GO Tau and DL Tau will result in significant excess emission in the joint regions. We thus perform a two-Gaussian fitting in the radial range covered by both components, which provides narrower ring width especially for DL Tau. In the illustration for a zoom-in view of the ring regions (Figure~\ref{fig:ring_image}), we adopt the two-Gaussian fitting results for GO Tau and DL Tau.   
Ring width measured from the image plane should always be taken as upper limits, as longer baseline data are under-weighted in CLEAN process. Therefore, we see that deconvolved ring width is always wider than that derived from visibility fitting for individual disks. 
Under the caveats, the behavior of a slightly narrower ring at 2.9\,mm is consistently seen for the dust ring in DS Tau. This pattern, a narrower dust ring at longer wavelength, is also observed for R73 ring in GO Tau, which is the second clearly detected ring in our sample. In addition, the large excess emission above simple Gaussian rings in DL Tau emphasizes that our derived dust ring parameters from parametric fitting in disks with such complex morphology probably suffer from large systematic uncertainties. 

\begin{deluxetable*}{lcccccccc}
\tabletypesize{\scriptsize}
\tablecaption{Dust ring properties from image plane fitting \label{tab:im_ring_prop}}
\tablewidth{0pt}
\tablehead{
\colhead{Name} & \colhead{Ring} & \colhead{Band} &  \colhead{$\sigma_{beam}$} & \colhead{Range} &  \multicolumn{3}{c}{Image Fitting}\\ 
\cline{6-9}
\colhead{} & \colhead{} & \colhead{} & \colhead{} & \colhead{} &  \colhead{R$_{im}$} & \colhead{$\sigma_{im}$} & \colhead{$\sigma_{de}$} & \colhead{$\sigma_{de}/H_{p}$} \\
\colhead{} & \colhead{} & \colhead{} & \colhead{(au)} & \colhead{(au)} & \colhead{(au)} & \colhead{(au)} & \colhead{(au)} \\
} 
\colnumbers
\startdata
DS Tau  & R57 &  B6 &  7.43 &  50-80   & 55.51   &  13.55  &  11.33  &  2.94  \\
  &  &  B3 &  7.43 &  50-80   & 54.80   &  13.15  &  10.85  &  3.03  \\
\hline
GO Tau  & R73 &  B6 &  7.03 &  65-80   & 72.27   &  12.38  &  10.19  &  1.56  \\
        &    &     &       &  65-130\tablenotemark{*}  & 71.42   &  11.56  &  9.18  &  1.41  \\
        &    &  B3 &  7.03 &  65-80   & 71.87   &  11.35  &  8.92   &  1.36  \\
        &    &     &       &  65-130\tablenotemark{*}  & 71.27   &  10.42  &  7.69  &  1.18  \\
GO Tau  & R110 &  B6 &  7.03 & 105-130  & 110.39  &  14.16  &  12.29  &  1.09  \\
        &    &     &       &  65-130\tablenotemark{*}  & 110.12   &  14.62  &  12.81 &  1.18  \\
        &    &  B3 &  7.03 & 105-130  & 111.51  &  14.31  &  12.46  &  1.11  \\
        &    &     &       &  65-130\tablenotemark{*}  & 111.25   &  14.74  &  12.95 &  1.15  \\
\hline
DL Tau  & R77 &  B6 &  7.60 &  70-90   & 77.05   &  24.16  &  22.93  &  4.69  \\
        &    &     &       &  70-130\tablenotemark{*}  & 74.81   &  17.77  &  16.06  &  3.48  \\
        &    &  B3 &  7.60 &  70-90   & 72.37   &  29.74  &  28.75  &  6.22  \\
        &    &     &       &  70-130\tablenotemark{*}  & 71.50   &  21.87  &  20.51  &  4.70  \\
DL Tau  & R116 &  B6 &  7.60 & 105-130  & 113.14  &  17.78  &  16.07  &  2.02 \\
        &    &     &       &  70-130\tablenotemark{*}  & 116.33   & 14.93  &  12.85  &  2.78  \\
        &    &  B3 &  7.60 & 110-130  & 111.76  &  17.52  &  15.79  &  2.06 \\
        &    &     &       &  70-130\tablenotemark{*}  & 116.02   & 14.88  &  12.79  &  2.93  \\
\enddata
\tablecomments{(1) Target name. (2) Dust ring name. (3) ALMA Band name. (4) Standard deviation of beam size. (5) Fitting range in radial profile. (6) Ring location from radial profile fitting. (7) Standard deviation of Gaussian ring from radial profile fitting. (8) Deconvolved standard deviation of Gaussian ring. (9) Ratio of $\sigma_{de}$ to the disk pressure scale height at ring location.  }
\tablenotetext{*}{Results from two-Gaussian fitting. }
\end{deluxetable*}

\end{CJK*}
\end{document}